\documentclass[onecolumn]{mn2e}
\newif\ifAMStwofonts
\def\pmb#1{\mbox{\boldmath$#1$}}
\def\gtsim {>\kern-1.2em\lower1.1ex\hbox{$\sim$}}
\def\ltsim {<\kern-1.2em\lower1.1ex\hbox{$\sim$}}
\def\gtsim {>\kern-1.2em\lower1.1ex\hbox{$\sim$}}
\def\ltsim {<\kern-1.2em\lower1.1ex\hbox{$\sim$}}
\def\ref{\hangindent=1pc \hangafter=1 \noindent}

\def\be{\begin{equation}}
\def\ee{\end{equation}}

\def\pmbmt#1{\pmb{\sf #1}}

\usepackage{epsfig}
\begin{document}

%%\input epsf

%\begintopmatter  %  start the two spanning material

\title{Non-axisymmetric low frequency oscillations of rotating and magnetized neutron stars}
\author[U. Lee]{Umin Lee$^1$\thanks{E-mail: lee@astr.tohoku.ac.jp}
\\$^1$Astronomical Institute, Tohoku University, Sendai, Miyagi 980-8578, Japan}

\date{Typeset \today ; Received / Accepted}
\maketitle

% \acceptedline is to be defined at the Journals office and not
% by an author.
%\acceptedline{Accepted 1988 December 15. Received 1988 December 14;
%  in original form 1988 October 11}

\begin{abstract}
We investigate non-axisymmetric low frequency modes of a rotating and magnetized neutron star, assuming
that the star is threaded by a dipole magnetic field whose strength at the stellar surface, $B_0$,
is less than $\sim 10^{12}$G, and whose magnetic axis is aligned with the rotation axis.
For modal analysis, we use a neutron star model composed of a fluid ocean, a solid crust, and a fluid core, where
we treat the core as being non-magnetic assuming that the magnetic pressure is much smaller than the gas pressure in the core.
For non-axisymmetric modes, 
spheroidal modes and toroidal modes are coupled in the presence of a magnetic field
even for a non-rotating star.
Here, we are interested in low frequency modes of a rotating and magnetized neutron star
whose oscillation frequencies are similar to
those of toroidal crust modes of low spherical harmonic degree and low radial order.
For a magnetic field of $B_0\sim 10^7$G, we find Alfv\'en waves 
in the ocean have similar frequencies to the toroidal crust modes, and
we find no $r$-modes confined in the ocean for this strength of the field.
We calculate the toroidal crustal modes, the interfacial modes 
peaking at the crust/core interface, and the core inertial modes and $r$-modes, and all these
modes are found to be insensitive to the magnetic field of strength $B_0\ltsim10^{12}$G.
We find the displacement vector of the core $l^\prime=|m|$ $r$-modes have large
amplitudes around the rotation axis at the stellar surface even in the presence of a surface magnetic 
field $B_0\sim10^{10}$G, where $l^\prime$ and $m$ are the spherical harmonic degree and the
azimuthal wave number of the $r$-modes, respectively.
We suggest that millisecond X-ray variations of accretion powered X-ray millisecond pulsars 
can be used as a probe into
the core $r$-modes destabilized by gravitational wave radiation.
If the $l^\prime=|m|=2$ $r$-mode is excited, we will have the pulsation 
of the frequency $\sim4\Omega/3$ with $\Omega$ being the spin frequency of the star.
\end{abstract}

\begin{keywords}
stars: neutron -- stars: oscillations -- stars : magnetic fields
\end{keywords}

\section{Introduction}

Oscillation of strongly magnetized neutron stars has attracted an intense attention in recent years, 
particularly motivated by
the discovery of quasi-periodic oscillations (QPOs) of magnetar candidates (e.g., Woods \& Thompson 2006), 
which are believed to be one of the observational manifestations of global oscillations 
of neutron stars that have a strong global magnetic field of order of $B_0\sim 10^{15}{\rm G}$
at the stellar surface
(e.g., Duncun 1998, Israel et al 2005; Strohmayer \& Watts 2005, 2006; Watts \& Strohmayer 2006).
Thus, recent theoretical studies of the oscillations of magnetized neutron stars have been
mainly concerned with the stars having extremely strong surface magnetic fields 
$B_0\gtsim 10^{15}$G, and these studies have suggested that
the QPOs observed in the magnetar candidates are attributable to the toroidal crust modes
and Alfv\'en modes of the neutron stars
(e.g., Piro 2005; Glampedakis, Samuelsson \& Andersson 2006; Lee 2007, 2008; 
Sotani, Kokkotas \& Stergioulas 2008;
Cerd\'a-Dur\'an, Stergioulas \& Font 2009; Colaiuda, Beyer \& Kokkotas 2009; Bastrukov et al 2009;
Sotani \& Kokkotas 2009).
For toroidal modes of strongly magnetized neutron stars, however,
it is intriguing, from the theoretical point of view, that Lee (2008) obtained
discrete frequency spectra of the magnetic modes, but Sotani, Kokkotas \& Stergioulas (2008), 
Cerd\'a-Dur\'an, Stergioulas \& Font (2009),
and Colaiuda, Beyer \& Kokkotas (2009), using a different numerical method from that used by Lee (2008), 
suggested the existence of continuum frequency spectra of the modes, as originally 
discussed by Levin (2006, 2007).

The burst oscillation observed in low mass X-ray binaries (LMXBs)
can be another example in which 
a magnetic field plays an important role in the oscillations of neutron stars, 
although the strength of the field at the surface of the neutron stars in LMXBs
is thought to be less than $\sim10^{10}$G, much weaker than that for the magnetar candidates.
For the burst oscillation, the
hot spot model (e.g., Strohmayer et al 1997; Cumming \& Bildsten 2001; Cumming et al 2002) and the
Rossby wave model (Heyl 2004; Lee 2004) have been proposed,
but it seems none of the models is accepted as the one that fully explains
the observational properties of the burst oscillation.
In the Rossby wave model, Heyl (2004, 2005) , Lee (2004), and 
Lee \& Strohmayer (2005) have examined the possibility that 
the burst oscillation is produced by low frequency Rossby waves, called $r$-modes in the astrophysical
literature, propagating in the surface fluid region (ocean), 
disregarding the effects of a magnetic field on the low frequency waves.
We note that Bildsten \& Cutler (1995) discussed the modal properties of $g$ modes propagating in the fluid ocean above the solid crust of mass accreting neutron stars as a possible mechanism
responsible for the $\sim 6$Hz QPOs observed in LMXBs.
In their paper, on the assumption that the magnetic pressure, $p_B$, 
is much smaller than the gas pressure, $p$,
the critical strength of a magnetic field, $B_c$, below which
the magnetic field has no significant effects on the $g$-modes,
was estimated to be $B_c\sim2\times 10^{10}$G for the accreted envelopes composed of carbon
(see also Piro \& Bildsten 2005).
Since their argument is based on a local analysis and on the assumption $p_B\ll p$, 
we think it useful to carry out a global modal analysis of low frequency waves 
propagating in the magnetized fluid ocean of a neutron star.

Accretion powered millisecond X-ray pulsars in LMXBs
show small amplitude, almost sinusoidal X-ray time variations,
the dominant period of which is thought to be the spin period of the underlying neutron stars.
Lamb et al (2009) argued that the millisecond X-ray variations
are produced by an X-ray emitting hot spot located at a magnetic pole rotating with 
the star, and that so long as the symmetry center of the hot spot is only slightly off the rotation axis
the X-ray variations produced have small amplitudes and become almost sinusoidal.
They also suggested that if the hot spot is located close to the rotation axis,
a slight drift of the hot spot away from the rotation axis leads to appreciable changes in the amplitudes and phases of the X-ray variations.
Lamb et al (2009) pointed out that a temporal
change in mass accretion rates and hence the radius of the magnetosphere, for example, can cause 
such a drift of the hot spot. 
We think it is also interesting to consider the possibility that
global oscillations of the neutron stars work as a mechanism that perturbs
the hot spot periodically.

In this paper, to calculate global oscillations of a rotating and magnetized neutron star,
we use the method of series expansion, in which the angular dependence of the perturbations
is represented by series expansion in terms of spherical harmonic functions with
different spherical harmonic degrees $l$ for a given azimuthal wave number $m$
(e.g., Lee \& Strohmayer 1996, Lee 2007).
We calculate low $m$, low frequency modes of a neutron star composed of a fluid ocean,
a solid crust, and a fluid core, where the star is assumed to be
threaded by a dipole magnetic field, the strength $B_0$ of 
which at the surface is smaller than $\sim10^{12}$G.
The method of calculation we employ is presented in \S 2, and
the numerical results are described in \S 3, 
and we discuss a local analysis for low frequency modes in the magnetized fluid ocean in \S 4, and we give discussion and conclusion in \S 5.

\section{Method of calculation}

We consider small amplitude oscillations of rotating and magnetized neutron stars in Newtonian dynamics, 
and no general relativistic effects on the oscillations are considered.
We employ spherical polar
coordinates $(r,\theta,\phi)$, whose origin is at the center of the star and the axis of rotation
is given by $\theta=0$.
We assume a dipole magnetic field given by
\be
\pmb{B}=\mu_m\nabla(\cos\theta/r^2),
\ee
where $\mu_m$ is the magnetic dipole moment.
For simplicity, we also assume that the magnetic axis coincides with the rotation axis.
Since the dipole field is a force-free field such that $(\nabla\times \pmb{B})\times\pmb{B}=0$, 
the field does not influence the equilibrium structure of the star.
When we assume the axis of rotation is also the magnetic axis,
the temporal and azimuthal angular dependence of perturbation can be represented by a single factor 
${\rm e}^{{\rm i}(m\phi+\omega t)}$,
where $m$ is the azimuthal wavenumber around the rotation axis and $\omega\equiv\sigma+m\Omega$ 
is the oscillation frequency observed in the corotating frame of the star, where
$\sigma$ is the oscillation frequency in an inertial frame and $\Omega$ is the angular frequency
of rotation.
The linearized basic equations used in a solid region of the star are given by
\be
-\omega^2\pmb{\xi}+2i\omega\pmb{\Omega}\times\pmb{\xi}=
{1\over\rho}\nabla\cdot\pmb{\sigma}^\prime
-{\rho^\prime\over\rho^2}\nabla\cdot\pmb{\sigma}
+{1\over4\pi\rho}\left(\nabla\times\pmb{B}^\prime\right)\times\pmb{B},
\ee
\be
\rho^\prime+\nabla\cdot(\rho\pmb{\xi})=0,
\ee
\be
{\rho^\prime\over\rho}={1\over\Gamma_1}{p^\prime\over p}-\xi_rA,
\ee
\be
\pmb{B}^\prime=\nabla\times(\pmb{\xi}\times\pmb{B}),
\ee
where $\rho$ is the mass density, $p$ is the pressure,
$\pmb{\xi}$ is the displacement vector, and $\rho^\prime$, $p^\prime$, and
$\pmb{B}^\prime$ are the Euler perturbations of the density, the pressure 
and the magnetic field, respectively,
and $A$ is the Schwartzshild discriminant defined by
\be
A={d\ln\rho\over dr}-{1\over\Gamma_1}{d\ln p\over dr},
\ee
and
\be
\Gamma_1=\left({\partial\ln p\over\partial\ln\rho}\right)_{ad}.
\ee
In equation (2), $\pmb{\sigma}^\prime$ denotes the Eulerian perturbation of the stress tensor, which
is derived by using the Lagrangian perturbation of the stress tensor defined,
in Cartesian coordinates, by
\be
\delta\sigma_{ij}=\left(\Gamma_1pu\right)\delta_{ij}
+2\mu\left(u_{ij}-{1\over 3}u\delta_{ij}\right),
\ee
where $u_{ij}$ is the strain tensor given by
\be
u_{ij}={1\over 2}\left({\partial\xi_i\over\partial x_j}+{\partial\xi_j\over\partial x_i}\right),
\ee
and $\mu$ is the shear modulus and $u=\Sigma_{l=1}^{3}u_{ll}$, and $\delta_{ij}$ is the Kronecker delta.
Note that we have employed the Cowling approximation neglecting 
the Eulerian perturbation of the gravitational potential, and that no effects of rotational
deformation are included.

We can obtain the equation of motion for a fluid region by replacing the terms $\pmb{\sigma}$
and $\pmb{\sigma}^\prime$ in equation (2) by $-p\delta_{ij}$ and $-p^\prime\delta_{ij}$,
respectively, and
we do not need to consider equations (8) and (9) for fluid regions.

Since the angular dependence of perturbations in a rotating and magnetized star cannot be represented by
a single spherical harmonic function, we expand the perturbed quantities in terms of 
spherical harmonic functions $Y_l^m\left(\theta,\phi\right)$ with different $l$s for a given $m$, 
assuming that the axis of rotation coincides with that of the magnetic field.
The displacement vector $\pmb{\xi}$ and the perturbed magnetic field $\pmb{B}^\prime$
are then represented by a finite series expansion of length $j_{\rm max}$ as
\be
{\pmb{\xi}\over r}=\sum_{j=1}^{j_{\rm max}}\left[S_{l_j}(r)Y^m_{l_j}(\theta,\phi)\pmb{e}_r
+H_{l_j}(r)\nabla_{\rm H} Y^m_{l_j}(\theta,\phi)
+T_{l^\prime_j}(r)~\pmb{e}_r\times\nabla_{\rm H} Y^m_{l^\prime_j}(\theta,\phi)\right]{\rm e}^{{\rm i}\omega t},
\ee
and 
\be
{\pmb{B}^\prime\over B_0(r)}=\sum_{j=1}^{j_{\rm max}}\left[b^S_{l^\prime_j}(r)Y^m_{l^\prime_j}(\theta,\phi)\pmb{e}_r
+b^H_{l^\prime_j}(r)\nabla_{\rm H} Y^m_{l^\prime_j}(\theta,\phi)
+b^T_{l_j}(r)~\pmb{e}_r\times\nabla_{\rm H} Y^m_{l_j}(\theta,\phi)\right]{\rm e}^{{\rm i}\omega t},
\ee
and the pressure perturbation, $p^\prime$, for example, is given by
\be
p^\prime=\sum_{j=1}^{j_{\rm max}}p^\prime_{l_j}(r)Y_{l_j}^m(\theta,\phi){\rm e}^{{\rm i}\omega t},
\ee
where
\be
\nabla_{\rm H}=\pmb{e}_\theta{\partial\over\partial\theta}+\pmb{e}_\phi{1\over\sin\theta}{\partial\over\partial\phi},
\ee
and $B_0(r)=\mu_m/r^3$,
and $l_j=|m|+2(j-1)$ and $l^\prime_j=l_j+1$ for even modes, and 
$l_j=|m|+2j-1$ and $l^\prime_j=l_j-1$ for odd modes, respectively, and $j=1,~2,~3,~\cdots, ~j_{\rm max}$.
Substituting the expansions (10), (11) and (12) into the linearized basic equations (2)$\sim$(5), 
and (8),
we obtain a finite set of coupled linear ordinary differential equations for the expansion coefficients
$S_{l_j}$, $H_{l_j}$, $T_{l^\prime_j}$ and so on, which we call
the oscillation equation.
For non-axisymmetric modes with $m\not=0$, we cannot expect
decoupling between spheroidal (polar) modes and
toroidal (axial) modes even for $\Omega=0$, although
for axisymmetric modes with $m=0$,
spheroidal modes and toroidal modes are decoupled when $\Omega=0$ (e.g., Lee 2007).

In this paper, we treat for simplicity the fluid core being non-magnetic, which may be justified
if we assume the magnetic pressure is much smaller than the gas pressure in the core.
The oscillation equations used in the magnetized regions are given in the Appendix, in which the jump conditions 
imposed at the solid/fluid interfaces and the boundary conditions applied at the stellar center 
and the surface are also given.
The oscillation equation used in the non-magnetic fluid core is found, for example, in
the Appendix of Lee \& Saio (1990).

\section{numerical results}

For the neutron star model used for modal analysis, we employ a cooling evolution model called NS05T7, which
was calculated by Richardson et al (1982).
The mass of the model is $M\sim0.5M_\odot$ and the central temperature is $T\sim10^7$K, and 
the model is composed of a fluid ocean, a solid crust and a fluid core, and 
because of density stratification $g$-modes propagate in the fluid regions.
The more detailed properties of the model, such as the equations of state used,
are described, for example, in McDermott, Van Horn \& Hansen (1988), 
who carried out modal analysis of the model assuming
no rotation and no magnetic fields.
Since this is a low mass model, the solid crust is rather thick, and the low radial order crustal modes
of low spherical harmonic degree $l$ are well separated from the $f$ and $p$ modes of the model,
which makes the modal analysis much simpler than the case in which the crustal modes have 
frequencies similar to those of the $f$ and $p$ modes as expected for more massive neutron stars
with accreted envelopes.

\begin{figure}
\resizebox{0.5\columnwidth}{!}{
\includegraphics{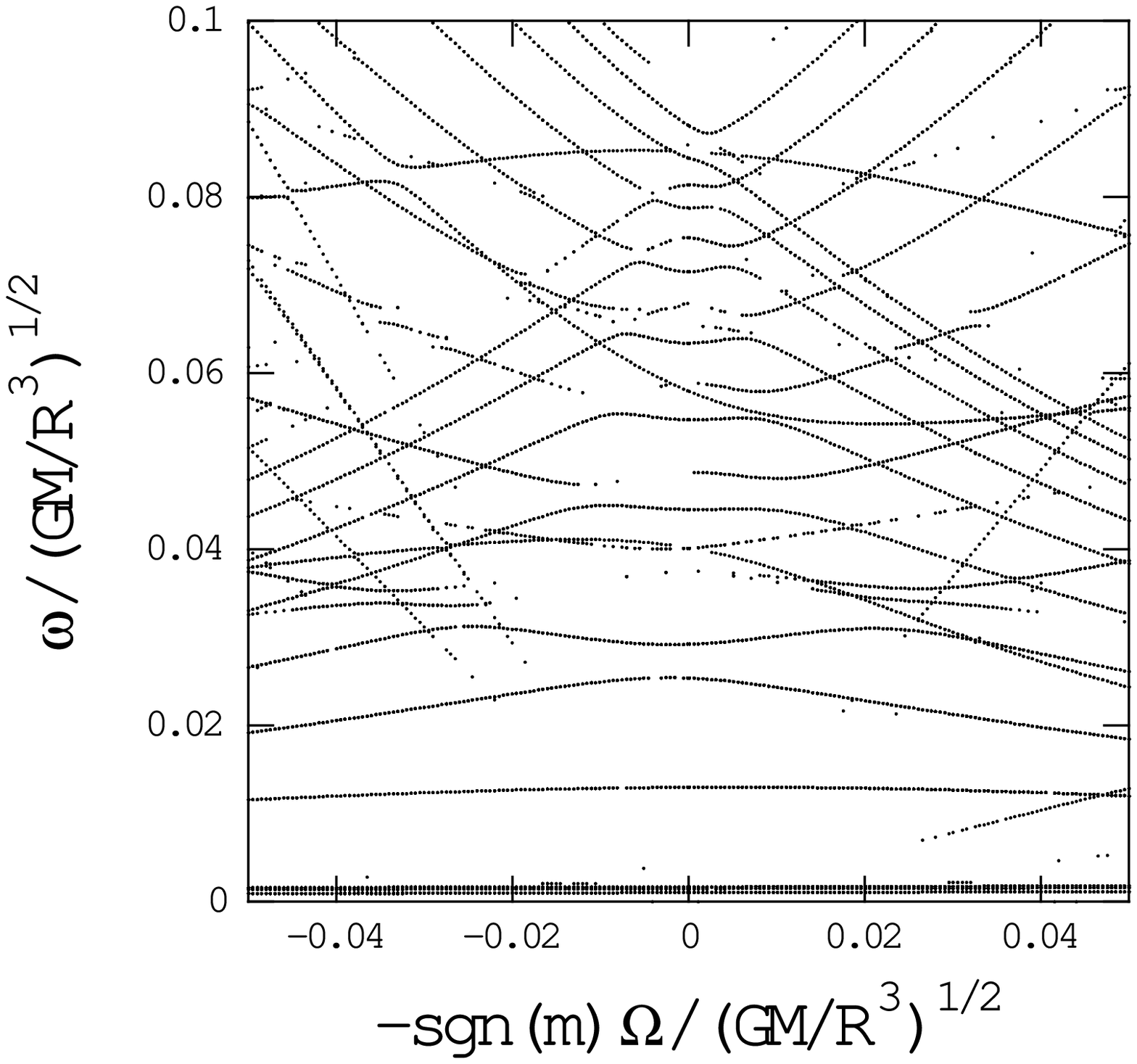}}
\resizebox{0.5\columnwidth}{!}{
\includegraphics{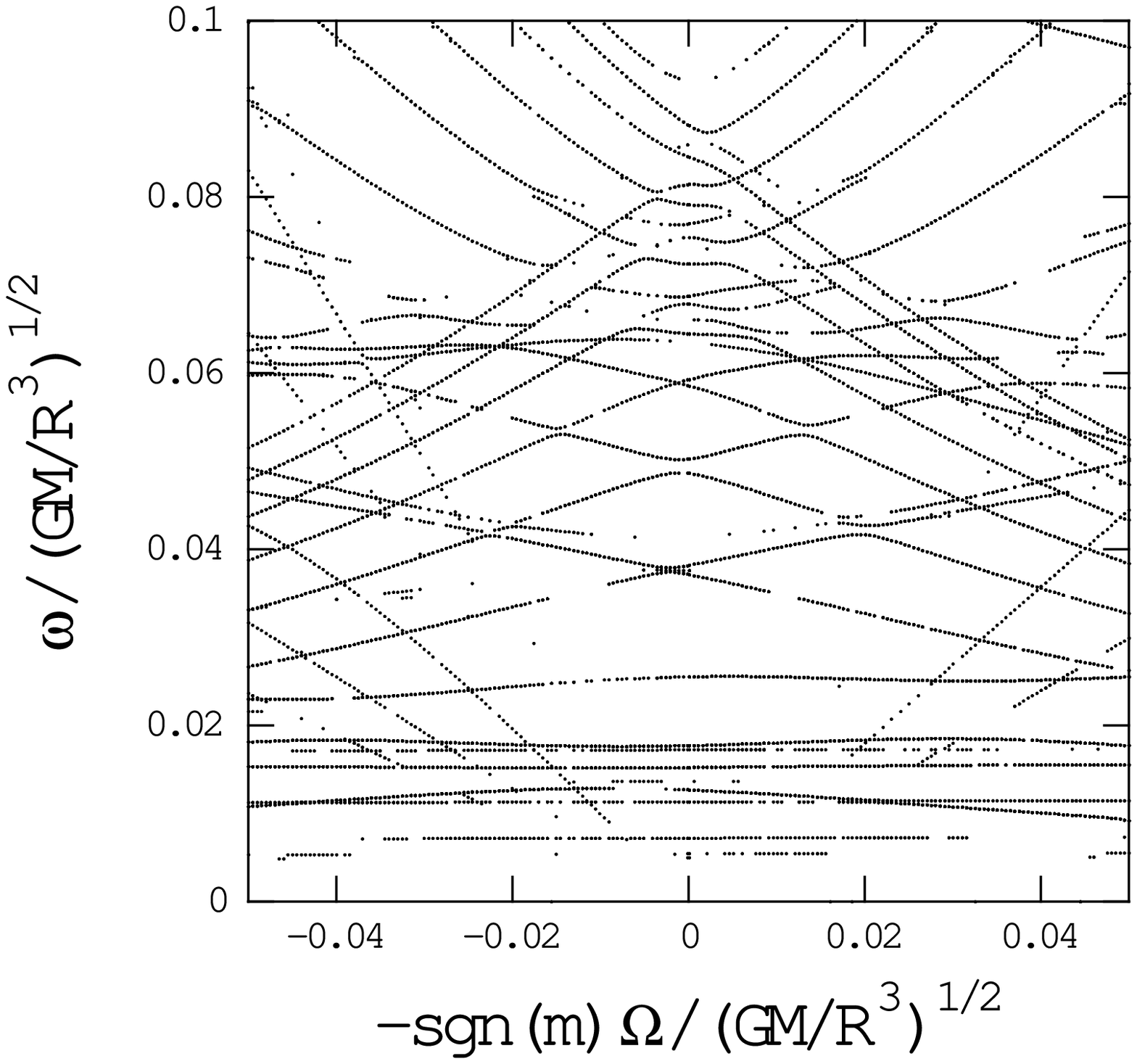}}
\caption{Low frequency $|m|=1$ modes of the model NS05T7 are plotted as functions of $-{\rm sgn}(m)\Omega$ 
for the case of $B_0=10^7$G for even modes in the left panel and odd modes in the right panel,
where ${\rm sgn}(m)=m/|m|$ and the positive and negative sides of the horizontal axis
are for prograde and retorgrade modes, respectively.
}
\end{figure}

Let us begin with the case of a weak magnetic field of strength $B_0=10^7$G.
In the presence of a magnetic field, the existence of Alfv\'en waves, whose
oscillation frequency depends on the strength of the magnetic field and 
on the direction of wave propagation relative to the magnetic field, 
inevitably makes frequency spectra complex.
For the case of $B_0=10^7$G, the frequency spectra of low frequency modes
are too complicated to calculate completely,
as suggested by 
Figure 1, which plots the oscillation frequency $\omega$ of low frequency $|m|=1$ modes
as functions of $\Omega$
for even modes (left panel) and odd modes (right panel), respectively, 
where we have used $j_{\rm max}=10$.
We notice that there appear in Figure 1 several kinds of modes, that is,
the inertial modes and $r$-modes in the fluid core, the toroidal crust modes, the interfacial modes
whose $\xi_r$ amplitudes peak at the interface between the core and the solid crust, and 
the modes confined in the fluid ocean.
The inertial modes and $r$-modes are rationally induced modes,
whose oscillation frequencies are approximately
proportional to $\Omega$ and are found on almost straight lines tending to the origin in the figure, 
where the $r$-modes appear on the side of retrograde modes.
The loci of the oscillation frequencies $\omega$ 
as functions of $\Omega$ for the core inertial modes, $r$-modes, the toroidal crust modes,
and the interfacial modes are more clearly seen for the case of a stronger magnetic field of
$B_0=10^{10}$G as shown by Figure 2 below.
We note that the toroidal crust modes are insensitive to the magnetic field of strength 
$B_0\ltsim 10^{12}$ (e.g., Lee 2007), and that the inertial modes and $r$-modes in 
the fluid core and the interfacial mode at the core/crust interface are not strongly affected by 
the magnetic field even if their eigenfunctions
extend to the surface through the magnetized solid crust and fluid ocean.

However, the modes that are confined in the surface ocean are 
strongly influenced by a magnetic field as weak as $B_0\sim10^7$G and show extremely complicated
frequency spectra.
Although 
there usually appear gravito-inertial modes and $r$-modes 
as non-axisymmetric low frequency modes confined in the ocean for a non-magnetized neutron star,
we find no ocean $r$-modes in the presence of a magnetic field, 
even if it is as weak as $B_0\sim10^7$G.
We also find that Alfv\'en modes 
come into the frequency spectra of the low frequency modes, modifying the spectra of the gravito-inertial modes.
We notice that there exist two branches of modes in Figure 1.
In one branch of modes, the oscillation frequency $\omega$ increases as $|\Omega|$ increases, while
the oscillation frequency decreases with increasing $|\Omega|$ in the other, 
although there occurs frequent avoided crossings between the two branches of modes as $\Omega$ varies.
Based on the local analysis presented in \S 4, we think the former can be regarded as 
(gravito-)inertial modes and the latter as Alfv\'en modes.
However, we have to note that the eigenfrequencies of the modes in the two branches do not necessarily reach
good convergence even if $j_{\rm max}$ is increased to 
$j_{\rm max}\sim20$.
In this sense we are not sure that the modes confined in the surface ocean we calculate
are discrete modes with real frequencies.

\begin{figure}
\resizebox{0.5\columnwidth}{!}{
\includegraphics{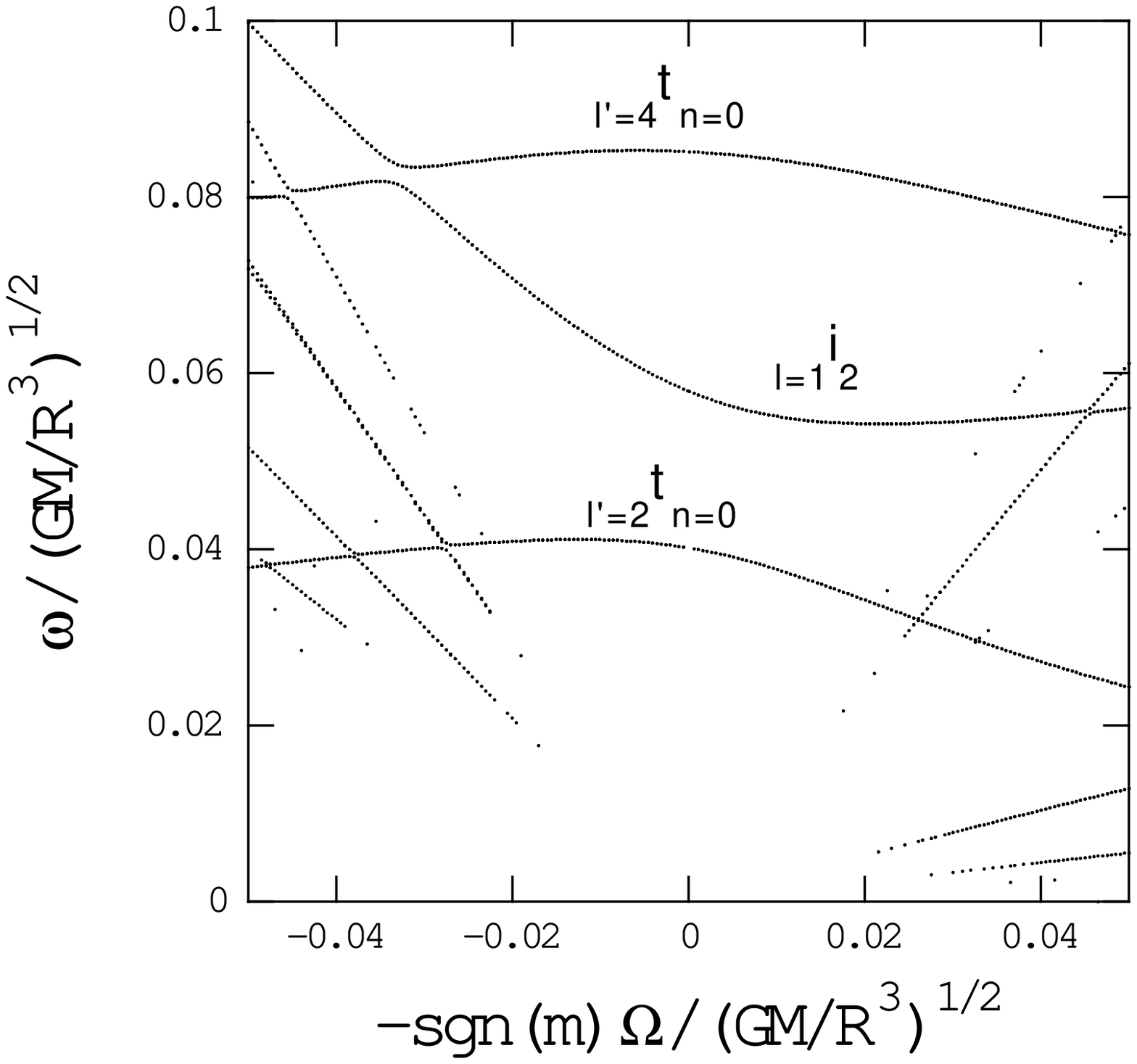}}
\resizebox{0.5\columnwidth}{!}{
\includegraphics{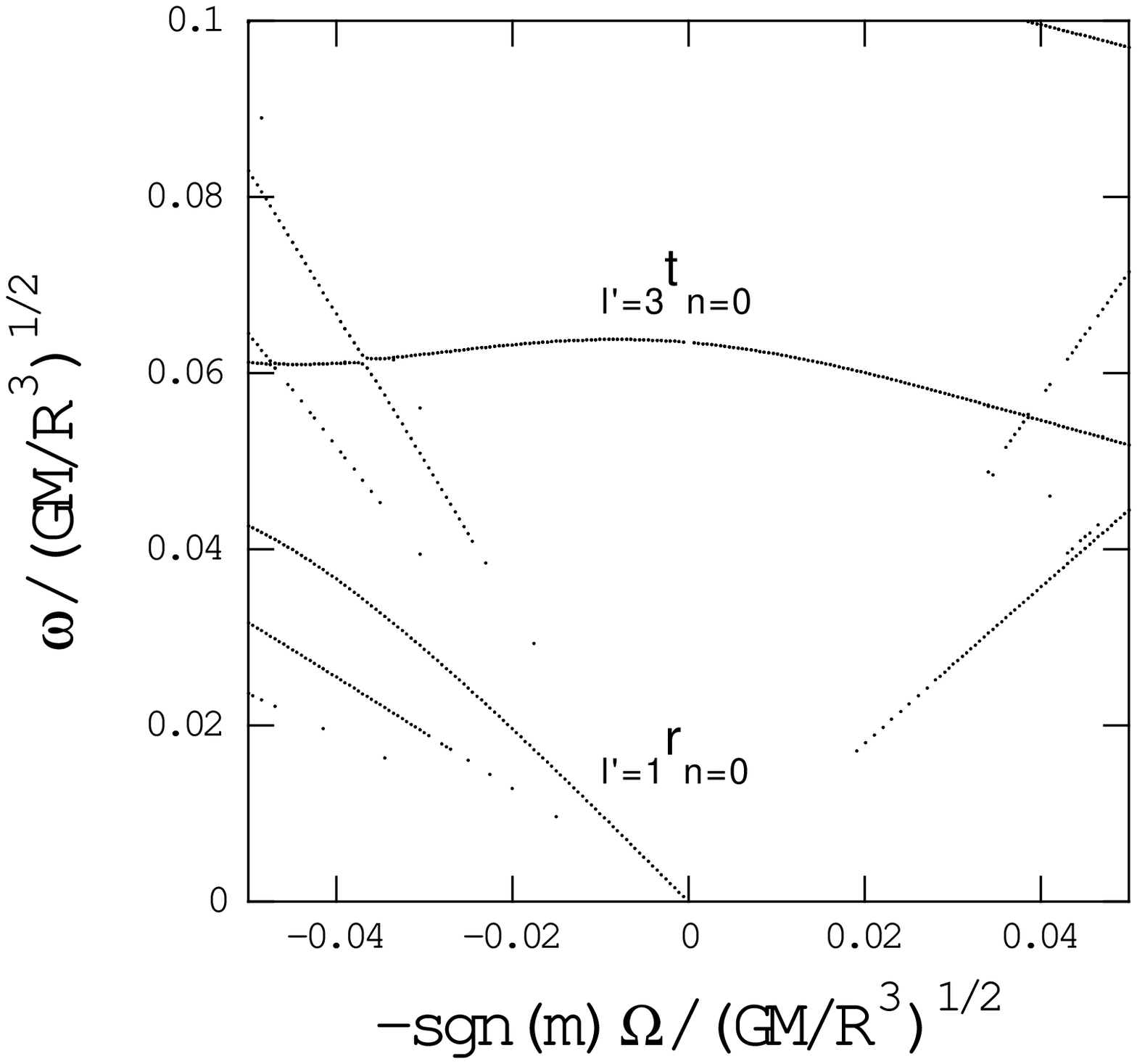}}
\caption{Low frequency $|m|=1$ modes of the model NS05T7 plotted versus
$-{\rm sgn}(m)\Omega$ for the case of $B_0=10^{10}$G for even modes in the left panel and
odd modes in the right panel, where ${\rm sgn(m)}=m/|m|$ and the
negative and positive sides of the horizontal axis are for retrograde and
prograde modes, respectively.
The symbols $_{l^\prime}t_n$, $_li_2$, and $_{l^\prime}r_n$ denote the toroidal crust modes,
the interfacial mode whose amplitudes peak at the core/crust interface, and the core $r$-mode,
respectively, where $l$ and $l^\prime$ denote the spherical harmonic degrees and the subscript $n$
is the radial order, which is usually equal to the number of radial nodes of the 
dominant component of the eigenfunctions.}
\end{figure}

In Figure 2, we plot low frequency $|m|=1$ modes against $\Omega$ 
for the case of $B_0=10^{10}$G for even modes (left panel)
and odd modes (right panel), where we have used $j_{\rm max}=10$.
In the figure,
the symbols $_{l^\prime}t_n$, $_li_2$, and $_{l^\prime}r_n$ respectively denote
the toroidal crust modes, the interfacial mode whose $\xi_r$ amplitudes peak
at the core/crust interface, and the core $r$-modes whose frequency $\omega$ tends to
$2m\Omega/\left[l^\prime\left(l^\prime+1\right)\right]$ as $\Omega\rightarrow 0$, where
$l$ and $l^\prime$ denote the spherical harmonic degrees
and the subscript $n$ indicates the radial order, usually corresponding to the number of nodes of the dominant eigenfunction in the propagation region.
Note that we do not attach any labels to inertial modes in the fluid core whose frequencies are
approximately proportional to $\Omega$ and are found on almost straight lines 
tending to the origin of the figure.
Because the fluid core is almost isentropic such that the Brunt-V\"ais\"al\"a frequency $N$
is extremely small, the odd $r$-mode labeled $_{l^\prime=|m|}r_{n=0}$ is the only $r$-mode we can find
for a given value of $m$ (e.g., Yoshida \& Lee 2000a).
We note that the eigenfrequencies of 
the modes plotted in Figure 2 obtain good convergence when $j_{\rm max}$ is increased to $j_{\rm max}\gtsim 10$.
We have calculated low frequency $|m|=1$ modes
for $B_0=10^{12}$G in the same frequency range as that in Figure 2, and obtained
the result quite similar to that for the case of $B_0=10^{10}$G.

Because of the effects of rotation,  
the frequencies $\omega$ of the toroidal crust modes and the interfacial mode
vary as $\Omega$ changes.
If we expand the oscillation frequency of a mode as $\omega=\omega_0+mC_1\Omega+O(\Omega^2)$, 
where $\omega_0$ is the frequency of the mode for $\Omega=0$, the $C_1$ coefficient
describes the first order response of the oscillation frequency to small $\Omega$, and
a list of the $C_1$ coefficients of $m=2$ spheroidal modes of the model NS05T7 is 
tabulated in Lee \& Strohmayer (1996), where no magnetic effects are considered.
For the toroidal crustal modes, on the other hand,
Lee \& Strohmayer (1996) showed $C_1=1/\left[l^\prime\left(l^\prime+1\right)\right]$
for non-magnetized stars.
Note that for rotationally induced modes such as inertial modes and $r$-modes, we have $\omega_0=0$.
Figure 2 indicates that the frequency
behavior of the crust modes for slow rotation is consistent to that expected from 
the $C_1$ coefficients, even in the presence of a magnetic field of strength
$B_0\sim10^{10}$G, although it is also clear that as $|\Omega|$ increases from $\Omega=0$,
the deviation of $\omega$ from the expansion $\omega_0+mC_1\Omega$ quickly becomes significant.
The deviation of the frequency from the expansion
may be partly caused by avoided crossings between two different modes.
An example found in Figure 2 is 
the avoided crossing between $_{l^\prime=2}t_{n=0}$ and $_{l=1}i_2$, which is
a crossing between a toroidal mode and a spheroidal mode.

In Figure 3, we plot the eigenfunctions $xS_{l_1}(x)$, $xH_{l_1}(x)$, and $xiT_{l^\prime_1}(x)$ 
of several low frequency $|m|=1$ 
modes for the case of $B_0=10^{10}$G and $\Omega/\sqrt{GM/R^3}=0.04$, 
where $x\equiv r/R$, and the low frequency modes we plot are the even toroidal crust mode
$_{l^\prime=2}t_{n=0}$, the even interfacial mode $_{l=1}i_2$, the odd $r$-mode $_{l^\prime=1}r_{n=0}$,
and the odd toroidal crustal mode $_{l^\prime=3}t_{n=0}$, and 
except for the $r$-mode, which is a retrograde mode with $m=1$,
all the other modes are prograde modes with $m=-1$.
Note that the toroidal crustal modes have appreciable amplitudes in the fluid core due to the
effects of rotation.
We find the eigenfunctions of the modes in the presence of
the dipole magnetic field of $B_0=10^{10}$G are quite
similar to those found in the absence of the magnetic field
(see, e.g., Lee \& Strohmayer 1996, Yoshida \& Lee 2001),
suggesting that these modes are not strongly affected by magnetic fields of that strength.
Note that in the vicinity of the stellar center, since the eigenfunctions $S_{l_1}$ and $H_{l_1}$ are
approximately proportional to $x^{l_1-2}$ (e.g., Unno et al 1989), 
the functions $xS_{l_1}$ and $xH_{l_1}$ behave as $\propto x^0$
for even modes of $l_1=|m|=1$ and as $\propto x^1$ for odd modes of $l_1=|m|+1=2$, as indicated by the panels (a) and (b).

\begin{figure}
\resizebox{0.33\columnwidth}{!}{
\includegraphics{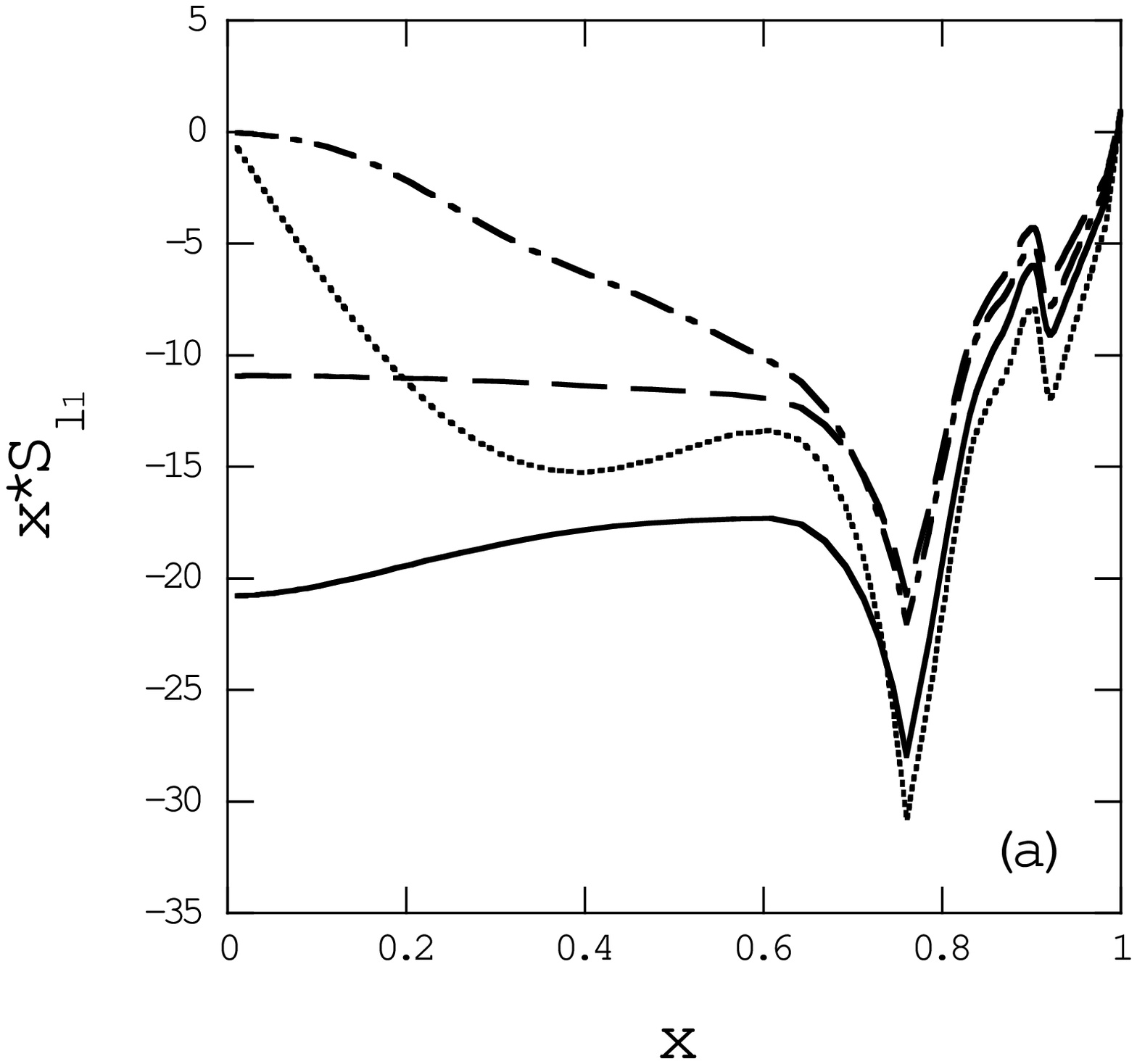}}
\resizebox{0.33\columnwidth}{!}{
\includegraphics{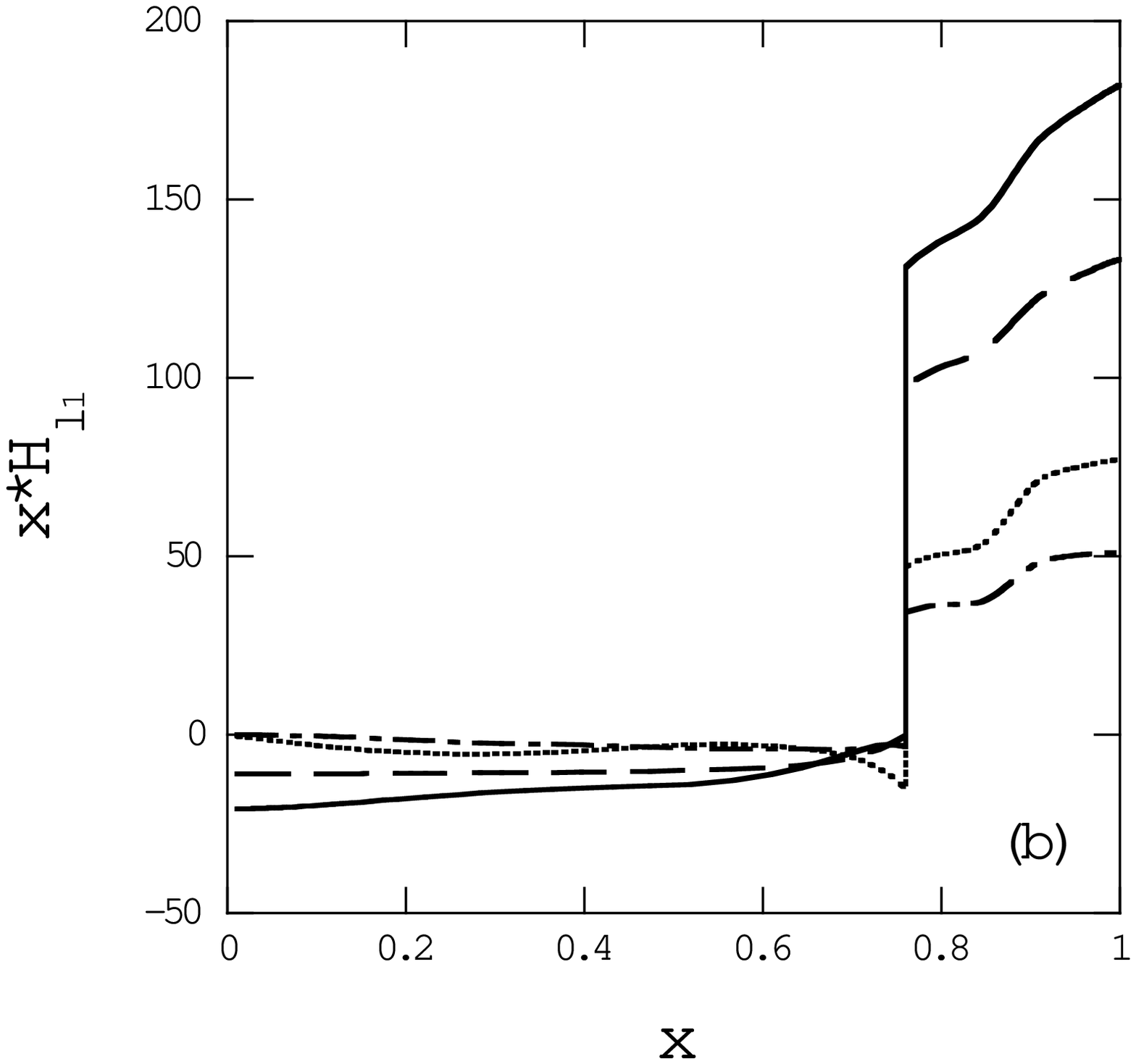}}
\resizebox{0.33\columnwidth}{!}{
\includegraphics{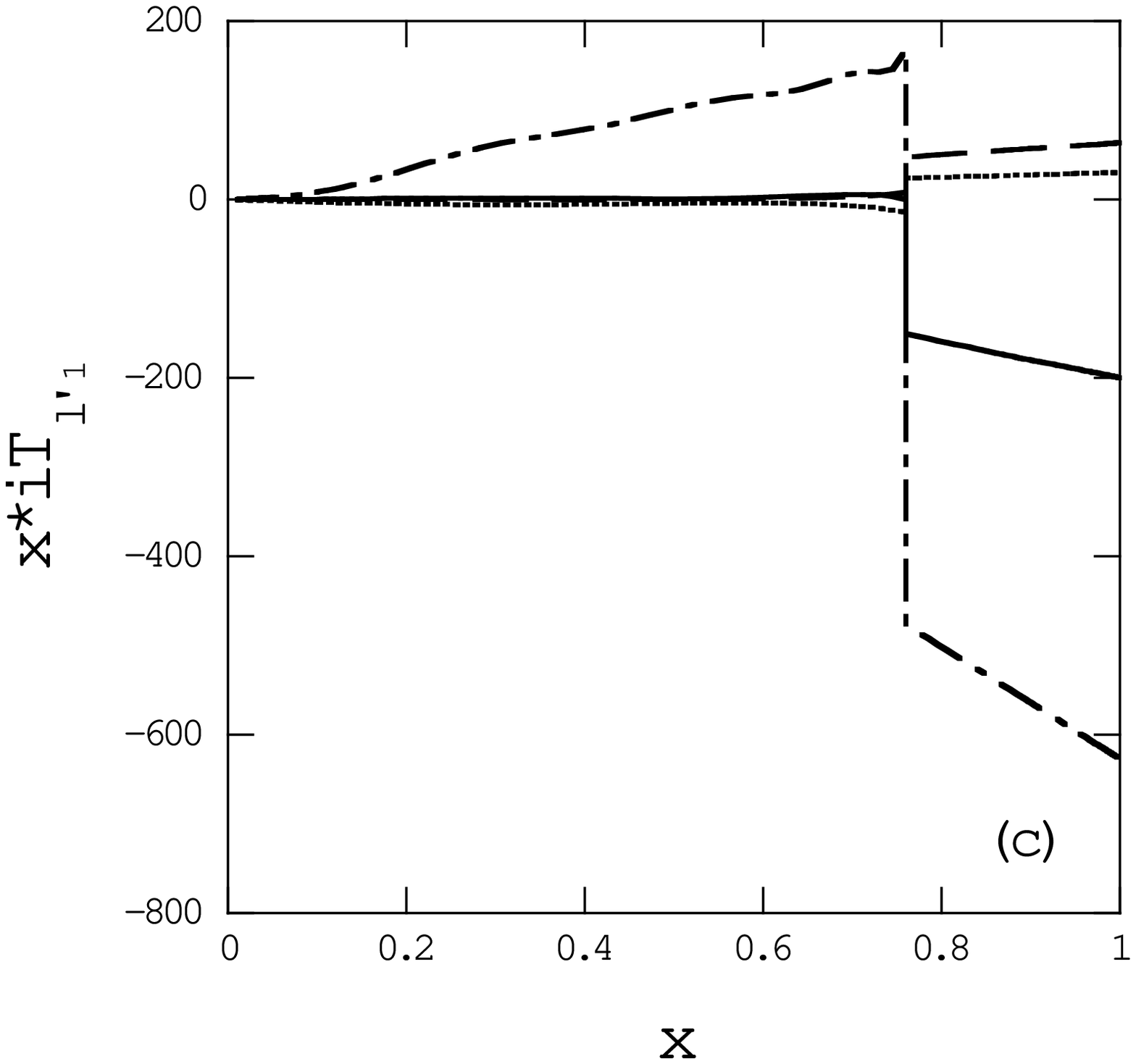}}
\caption{Eigenfunctions $xS_{l_1}$ (panel a), $xH_{l_1}$ (panel b), and 
$xiT_{l^\prime_1}$ (panel c) plotted
against $x\equiv r/R$ for the low frequency $|m|=1$ modes for the case of $B_0=10^{10}$G and
$\Omega/\sqrt{GM/R^3}=0.04$, where we use $j_{\rm max}=10$.
In each panel, the solid line is for the even toroidal crustal mode $_{l^\prime=2}t_{n=0}$, 
the dashed line for the even interfacial mode $_{l=1}i_2$, 
the dash-dotted line for the odd $r$-mode $_{l^\prime=1}r_{n=0}$, and 
the dotted line for the odd toroidal crustal mode $_{l^\prime=3}t_{n=0}$.
Note that the amplitude normalization is given by $xS_{l_1}=1$ at the stellar surface.
Except for the $r$-mode, which is a retrograde mode with $m=1$, all the other modes are
prograde modes with $m=-1$.
}
\end{figure}

From the observational point of view, it is useful to know the $\theta$ dependence of 
the displacement vector of the modes at the surface of the star.
As noted in the previous section, for rotating and magnetized stars,
the eigenfunction of an oscillation mode cannot be represented by a single
spherical harmonic function $Y_l^m\left(\theta,\phi\right)$ and hence their surface pattern
can be largely different from that for non-magnetic and non-rotating stars.
To see the angular dependence of the displacement vector at the surface,
we introduce the functions $X^j\left(\theta\right)$ defined by
\be
X^r\left(\theta\right)e^{{\rm i}m\phi}=\sum _{j=1}^{j_{\rm max}}S_{l_j}\left(R\right)Y_{l_j}^m\left(\theta,\phi\right)
\ee
\be
X^\theta\left(\theta\right)e^{{\rm i}m\phi}=\pmb{e}_\theta\cdot\sum_{j=1}^{j_{\rm max}}\left[
H_{l_j}(R)\nabla_{\rm H} Y^m_{l_j}(\theta,\phi)
+T_{l^\prime_j}(R)~\pmb{e}_r\times\nabla_{\rm H} Y^m_{l^\prime_j}(\theta,\phi)\right],
\ee
\be
X^\phi\left(\theta\right)e^{{\rm i}m\phi}=-{\rm i}\pmb{e}_\phi\cdot\sum_{j=1}^{j_{\rm max}}\left[
H_{l_j}(R)\nabla_{\rm H} Y^m_{l_j}(\theta,\phi)
+T_{l^\prime_j}(R)~\pmb{e}_r\times\nabla_{\rm H} Y^m_{l^\prime_j}(\theta,\phi)\right].
\ee
In Figure 4, we plot the functions $X^j\left(\theta\right)$ of the low frequency modes
presented in Figure 3 for the case of $B_0=10^{10}$G and $\Omega/\sqrt{GM/R^3}=0.04$, 
where 
in each of the panels the solid, dashed, and dotted lines in each panel respectively denote the functions
$X^r$, $X^\theta$, and $X^\phi$, which are normalized by their maximum amplitudes.
Since the modes have low frequencies, the maximum amplitudes of $\xi_r$ are much smaller than
those of $\xi_\theta$ and $\xi_\phi$, that is, the horizontal and/or toroidal components
of the displacement vector are dominant over the radial component.
It is to be noted that although the
amplitudes of the ocean $r$-modes tend to be confined to the equatorial regions (see, e.g., Lee 2004), 
this is not the case
for the $l^\prime=|m|$ $r$-modes in the fluid core, which have large amplitudes in the regions around 
the the poles.

\begin{figure}
\resizebox{0.8\columnwidth}{!}{
\includegraphics{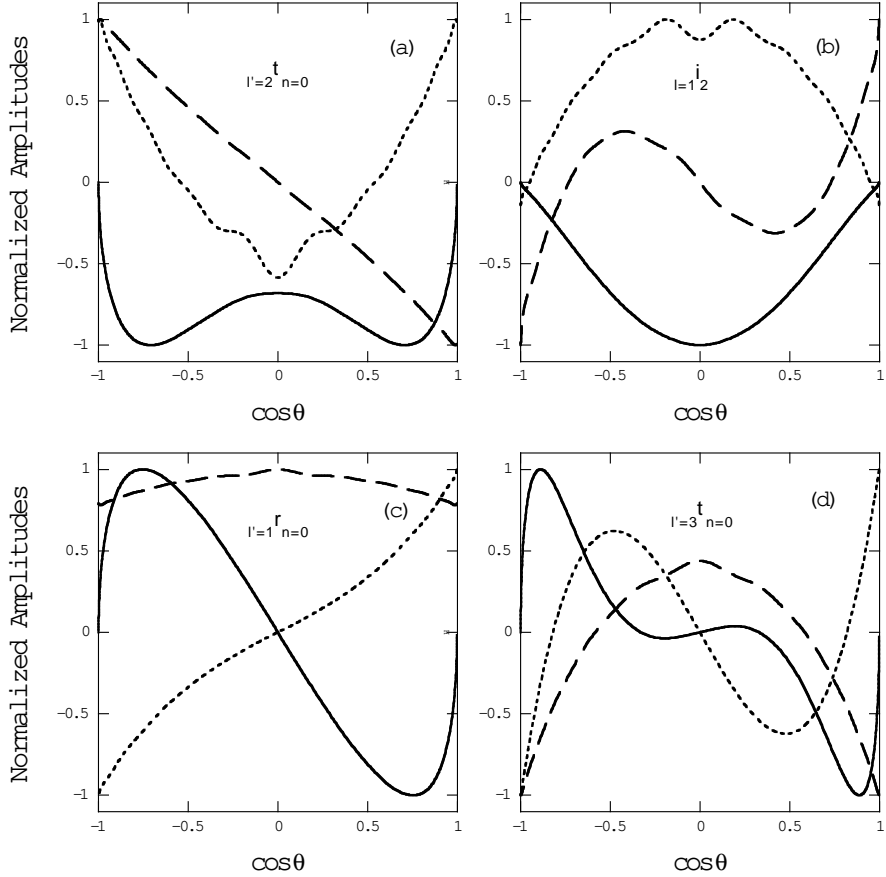}}
\caption{Functions $X^r(\theta)$, $X^\theta(\theta)$, and $X^\phi(\theta)$ versus $\cos\theta$ for the low 
frequency $|m|=1$ modes for the case of $B_0=10^{10}$G and $\Omega/\sqrt{GM/R^3}=0.04$ for $j_{\rm max}=10$, 
where the functions are evaluated at the stellar surface, and
the solid, dashed, and dotted lines denote
the functions $X^r(\theta)$, $X^\theta(\theta)$, and $X^\phi(\theta)$, respectively.
Panels (a), (b), (c), and (d) are for the even toroidal crust mode $_{l^\prime=2}t_{n=0}$, 
the even interfacial mode $_{l=1}i_2$,
the odd core $_{l^\prime=1}r_{n=0}$ mode, and the odd toroidal crust mode $_{l^\prime=3}t_{n=0}$, respectively, 
where 
only the $r$-mode is a retrograde mode with $m=1$, and the others are prograde modes with $m=-1$.
The functions are normalized by their maximum amplitudes.
The functions $X^r(\theta)$ and $X^\phi(\theta)$ of even (odd) modes are symmetric (antisymmetric)
with respect to the axis of $\cos\theta=0$, while the function $X^\theta(\theta)$ of even (odd) modes is antisymmetric (symmetric).
}
\end{figure}

In Figure 5, we plot the frequencies $\omega$ of low frequency $|m|=2$ modes against $\Omega$ 
for the case of $B_0=10^{10}$G for even modes (left panel)
and odd modes (right panel), where we have used $j_{\rm max}=10$.
For the case of $|m|=2$, the toroidal crustal mode $_{l^\prime=3}t_{n=0}$ appears as an even mode,
while $_{l^\prime=2}t_{n=0}$ and $_{l^\prime=4}t_{n=0}$ as an odd mode.
We note that no interfacial modes appear in the frequency range shown in the figure.
Figure 6 shows the functions $X^j(\theta)$ versus $\cos\theta$ for the core $l^\prime=m=2$ $r$-mode
at $\Omega/\sqrt{GM/R^3}=0.04$
for $B_0=10^{10}$G (left panel) and $B_0=0$G (right panel), where the functions, evaluated at the
stellar surface, are normalized by their maximum amplitudes.
This figure shows that
the surface pattern generated by the core $l^\prime=m=2$ $r$-mode is not affected 
by the field as strong as $B_0=10^{10}$G, which is also the case for the core $l^\prime=m=1$ $r$-mode.

\begin{figure}
\resizebox{0.5\columnwidth}{!}{
\includegraphics{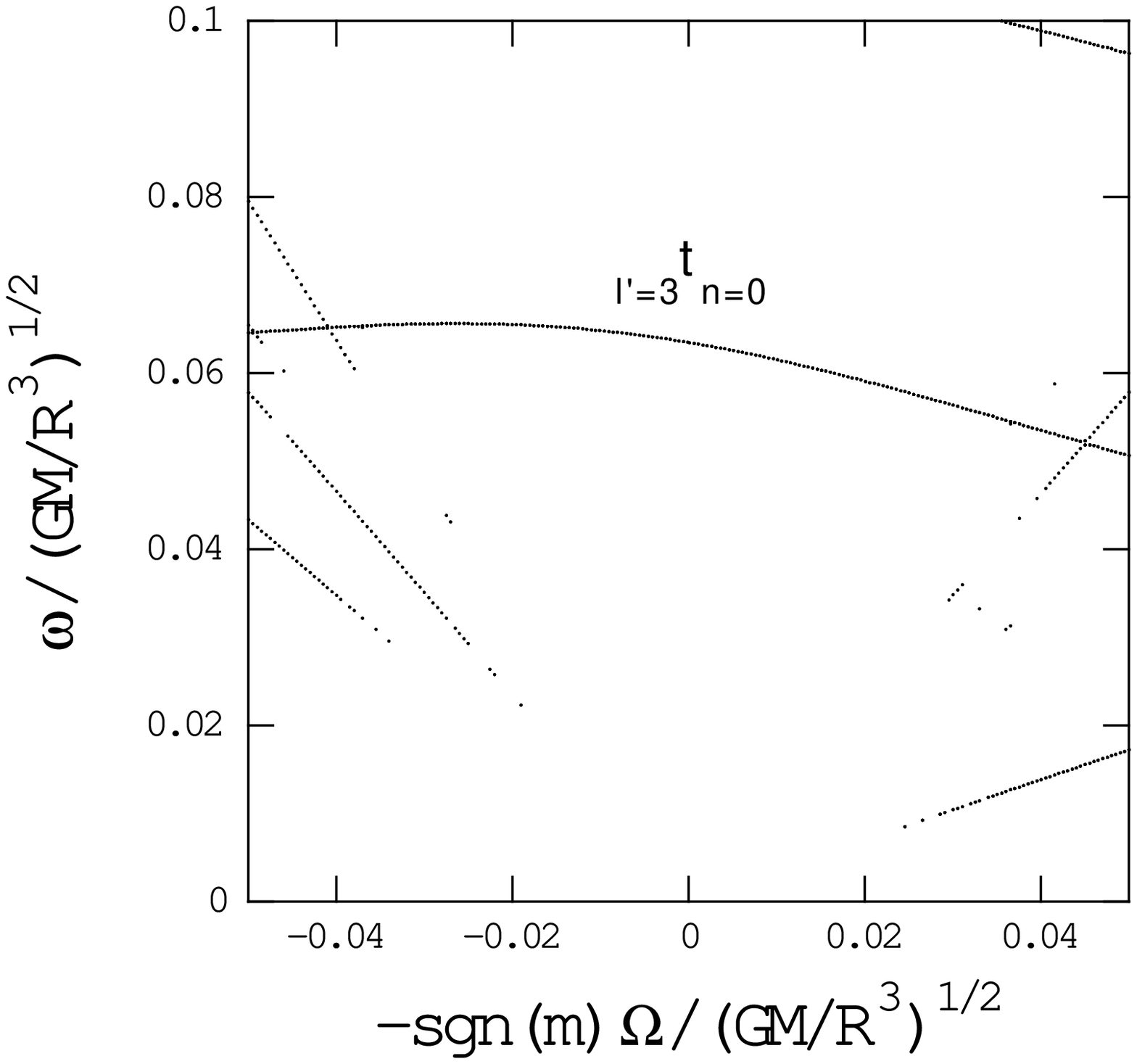}}
\resizebox{0.5\columnwidth}{!}{
\includegraphics{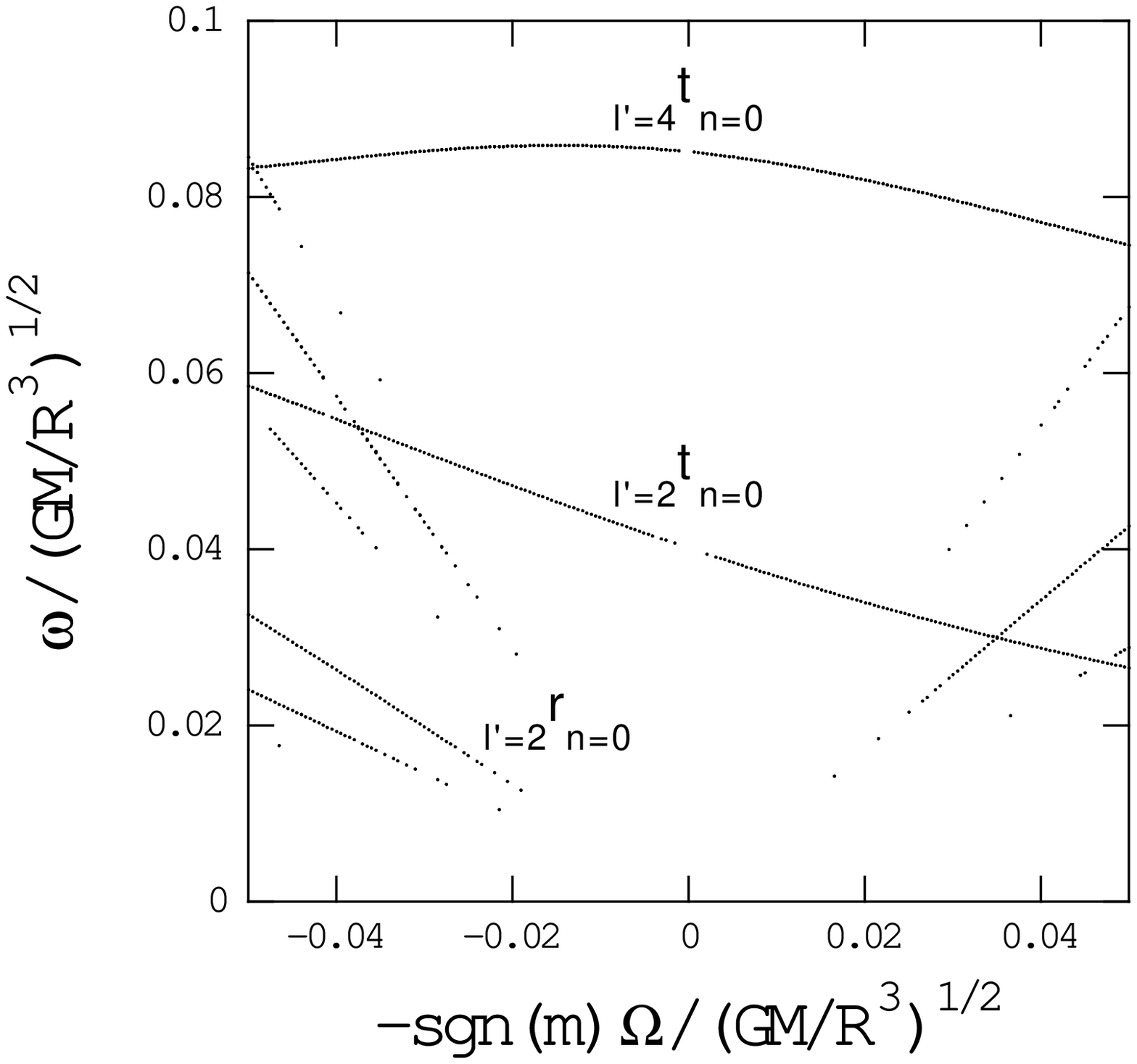}}
\caption{Same as Figure 2 but for $|m|=2$.}
\end{figure}

\begin{figure}
\resizebox{0.5\columnwidth}{!}{
\includegraphics{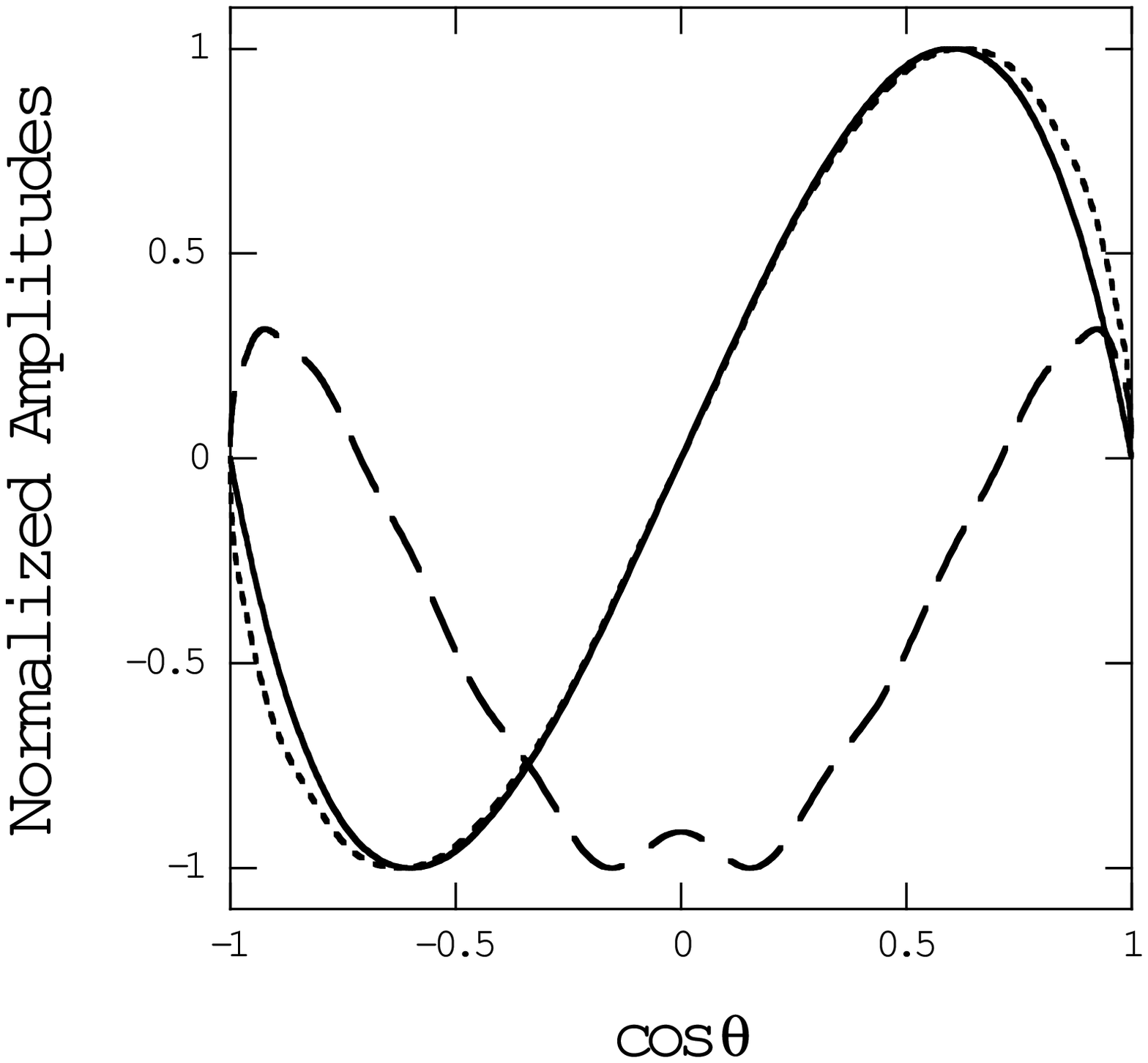}}
\resizebox{0.5\columnwidth}{!}{
\includegraphics{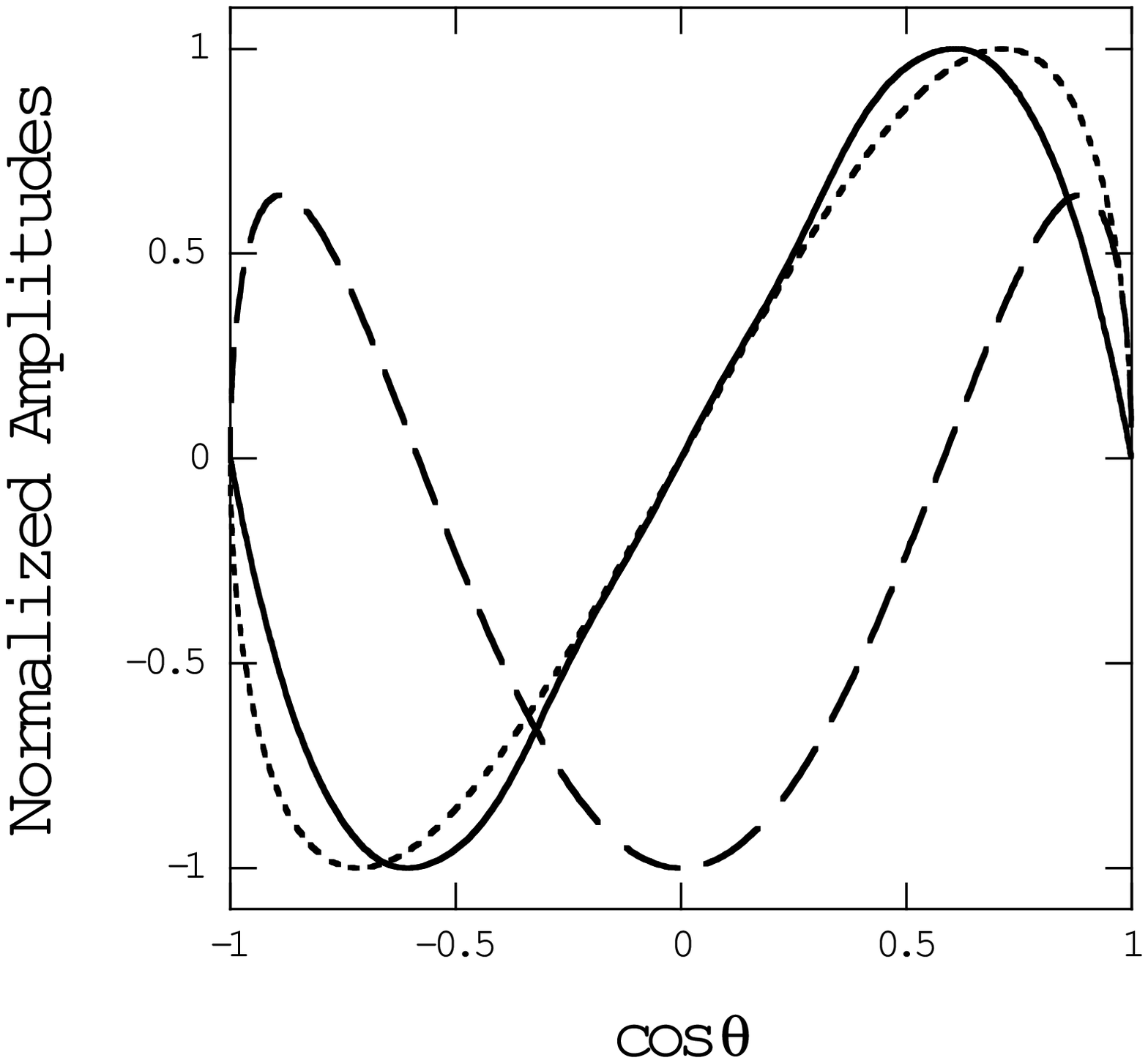}}
\caption{Functions $X^r(\theta)$, $X^\theta(\theta)$, and $X^\phi(\theta)$ at the surface
versus $\cos\theta$ for the $m=2$ $_{l^\prime=2}r_{n=0}$ mode for the cases of $B_0=10^{10}$G 
in the left panel and $B_0=0$G in the right panel, 
where $\Omega/\sqrt{GM/R^3}=0.04$ and $j_{\rm max}=10$ are assumed.
The solid, dashed, and dotted lines respectively denote
the functions $X^r(\theta)$, $X^\theta(\theta)$, and $X^\phi(\theta)$, 
which are normalized by their maximum amplitudes.
}
\end{figure}

\section{Local Analysis}

To gain an understanding of the low frequency modes confined in the shallow fluid ocean 
calculated for the case of $B_0=10^7$G, we apply a local analysis to low frequency modes
in a fluid region of a rotating and magnetized neutron star.
Note that a local analysis of waves in rotating stars with no magnetic fields is
found in, for example, Unno et al (1989).
In this section,
we assume that the $\pmb{x}$ and $t$ dependence of the perturbed quantities is
given by the function $\exp\left[{{\rm i}\left(\pmb{k}\cdot\pmb{x}+ \omega t\right)}\right]$,
where $\pmb{k}$ is the wavenumber vector.
Using the equation of motion for a fluid region and equations (3) to (5), 
we can derive a set of equations used for the local analysis:
\be
-\omega^2\pmb{\xi}+2i\omega\pmb{\Omega}\times\pmb{\xi}=-{1\over\rho}{\rm i}\pmb{k} p^\prime
+{\rho^\prime\over\rho^2}\nabla p+{1\over 4\pi\rho}\left({\rm i}\pmb{k}\times\pmb{B}^\prime\right)\times\pmb{B},
\ee
\be
\rho^\prime+{\rm i}\rho\left(\pmb{k}\cdot\pmb{\xi}\right)+\pmb{\xi}\cdot\nabla\rho=0,
\ee
\be
p^\prime=\Gamma_1p\left(\pmb{\xi}\cdot\pmb{A}+\rho^\prime/\rho\right),
\ee
\be
\pmb{B}^\prime={\rm i}\pmb{k}\times\left(\pmb{\xi}\times\pmb{B}\right),
\ee
which are combined to give
\be
\left[\omega^2-{\left(\pmb{k}\cdot\pmb{B}\right)^2\over 4\pi\rho}\right]\pmb{\xi}
-\left[\left(a^2+a_A^2\right)\pmb{k}-{\left(\pmb{k}\cdot\pmb{B}\right)\over 4\pi\rho}\pmb{B}
\right]\left(\pmb{k}\cdot\pmb{\xi}\right)
+{\left(\pmb{k}\cdot\pmb{B}\right)\left(\pmb{\xi}\cdot\pmb{B}\right)\over 4\pi\rho}\pmb{k}
-2{\rm i}\omega\pmb{\Omega}\times\pmb{\xi}-{\rm i}a^2\left(\pmb{\xi}\cdot\pmb{A}\right)\pmb{k}
-{\rm i}\left(\pmb{k}\cdot\pmb{\xi}\right){\nabla p\over \rho}=0,
\ee
where $a=\sqrt{\Gamma_1p/\rho}$ is the adiabatic sound speed and
$a_A=B/\sqrt{4\pi\rho}$ is the Alfv\'en velocity and $B=|\pmb{B}|$.

To make the local analysis tractable, we employ a local Cartesian coordinate system whose
$z$-axis is along the radial direction, and we assume that $\pmb{\Omega}=\Omega_z\pmb{e}_z$
neglecting the local horizontal component of the rotation vector.
Equation (21) can be rewritten into a form $\pmbmt{W}\pmb{\xi}=0$ with $\pmbmt{W}$ being a matrix, and
the condition $\det\pmbmt{W}=0$ leads to the dispersion relation:
\be
\omega^6+A_4\omega^4+A_2\omega^2+A_1\omega+A_0=0,
\ee
where
\be
A_4=-\left(a^2+a_A^2\right)k^2-a_A^2k^2\cos^2\theta-\left(2\Omega_z\right)^2
-\beta^2,
\ee
\begin{eqnarray}
A_2=\left(2a^2+a_A^2\right)a_A^2k^4\cos^2\theta+a^2k_H^2N^2+\left(2\Omega_z\right)^2\left[a^2k_z^2+a_A^2k^2\left(\cos^2\theta
+{k_z^2\over k^2}-2\cos\theta{B_z\over B}{k_z\over k}\right)\right] \nonumber\\
+\left[a_A^2k_H^2+2a_A^2k^2\cos\theta{B_z\over B}{k_z\over k}+\left(2\Omega_z\right)^2\right]\beta^2,
\end{eqnarray}
\be
A_1=-4a_A^2gk^2k_z\Omega_z{k_H\over k}{B_H\over B}\sin\psi\cos\theta,
\ee
\be
A_0=-a^2a_A^2k^4\cos^2\theta\left(a_A^2k^2\cos^2\theta+{k_H^2\over k^2}N^2\right)
-a_A^4k^4\cos^2\theta{B_z^2\over B^2}\beta^2,
\ee
where $g=GM_r/r^2$, $\cos\theta=\left(\pmb{k}\cdot\pmb{B}\right)/\left(kB\right)$, $k=|\pmb{k}|$,
$B_H=\sqrt{B_x^2+B_y^2}$,
$k_H=\sqrt{k_x^2+k_y^2}$, $\sin\psi=\left(\pmb{k}\times\pmb{B}\right)_z/\left(k_HB_H\right)$,
and $N=\sqrt{-gA}$ is the Brunt-V\"ais\"al\"a frequency, and
\be
\beta^2\equiv -{d\ln\rho\over dz}g=N^2+{g^2\over a^2}.
\ee
Note that the term proportional to $A_1$ breaks the symmetry given by 
$\omega\left(-\Omega_z\right)=\omega\left(\Omega_z\right)$.

If we assume $\pmb{B}=0$ and $\pmb{\Omega}\not=0$, the dispersion relation reduces to
\be
\omega^2\left\{\omega^4-\left[a^2k^2+\left(2\Omega_z\right)^2+\beta^2\right]\omega^2
+\left[a^2k_z^2\left(2\Omega_z\right)^2+a^2k_H^2N^2+\left(2\Omega_z\right)^2\beta^2\right]\right\}=0,
\ee
the non-trivial solution of which is
\be
\omega^2={1\over 2}\left\{a^2k^2+\left(2\Omega_z\right)^2+\beta^2
\pm\sqrt{
\left[a^2k^2+\beta^2-\left(2\Omega_z\right)^2\right]^2
-4a^2k_H^2\left[N^2-\left(2\Omega_z\right)^2\right]}\right\}.
\ee
On the other hand, if we assume $\pmb{\Omega}=0$ and $\pmb{B}\not=0$, 
the dispersion relation reduces to
\begin{eqnarray}
\left(\omega^2-a_A^2k^2\cos^2\theta\right)\left[\omega^4-\left(a^2k^2+a_A^2k^2+\beta^2\right)\omega^2
+a^2k^2\left(a_A^2k^2\cos^2\theta+{k_H^2\over k^2}N^2\right)
+a_A^2k^2\left({B_z^2\over B^2}+{k_H^2\over k^2}{B_H^2\over B^2}\sin^2\psi\right)\beta^2
\right] \nonumber \\
+a_A^4k^4{k_H^2\over k^2}{B_H^2\over B^2}\sin^2\psi\cos^2\theta\beta^2=0.
\end{eqnarray}
If we can further assume $\sin\psi=0$, the solutions of the dispersion relation are separated into
\be
\omega^2=a_A^2k^2\cos^2\theta,
\ee
corresponding to the Alfv\'en waves, and to
\be
\omega^2={1\over 2}\left\{\left(a^2+a_A^2\right)k^2+\beta^2\pm\sqrt{\left[\left(a^2+a_A^2\right)k^2+\beta^2\right]^2
-4a^2k^2\left(a_A^2k^2\cos^2\theta
+\displaystyle{ k_H^2\over k^2}N^2+{a_A^2\over a^2}{B_z^2\over B^2}\beta^2\right)}\right\}.
\ee

For the case of $\pmb{B}\not=0$ and $\pmb{\Omega}\not=0$,
it is difficult to analytically solve the dispersion relation (22) in general .
Here, we numerically solve equation (22), which can be rewritten,
by normalizing various quantities, into
\be
\bar\omega^6+{A_4\over \Omega_0^2}\bar\omega^4+{A_2\over\Omega_0^4}\bar\omega^2
+{A_1\over \Omega_0^5}\bar\omega+{A_0\over \Omega_0^6}=0,
\ee
where
\be
{A_4\over \Omega_0^2}=-\left(p+q\right)\left(Rk\right)^2-q\left(Rk\right)^2\cos^2\theta
-4\bar\Omega_z^2-{\beta^2\over\Omega_0^2},
\ee
\begin{eqnarray}
{A_2\over\Omega_0^4}=\left(2p+q\right)q\left(Rk\right)^4\cos^2\theta+p\left(Rk\right)^2{k_H^2\over k^2}\bar N^2
+4\bar\Omega_z^2\left[p\left(Rk\right)^2{k_z^2\over k^2}+q\left(Rk\right)^2
\left(\cos^2\theta+{k_z^2\over k^2}-2\cos\theta{B_z\over B}{k_z\over k}\right)\right] \nonumber \\
+\left[q\left(Rk\right)^2{k_H^2\over k^2}+2q\left(Rk\right)^2\cos\theta{B_z\over B}{k_z\over k}
+4\bar\Omega_z^2\right]{\beta^2\over\Omega_0^2},
\end{eqnarray}
\be
{A_1\over\Omega_0^5}=-4q{g\over g_S}\left(Rk\right)^3\bar\Omega_z{k_z\over k}{k_H\over k}{B_H\over B}\sin\psi\cos\theta,
\ee
\be
{A_0\over\Omega_0^6}=-pq\left(Rk\right)^4\cos^2\theta\left[q\left(Rk\right)^2\cos^2\theta
+{k_H^2\over k^2}\bar N^2\right]-q^2\left(Rk\right)^4\cos^2\theta
{B_z^2\over B^2}{\beta^2\over\Omega_0^2},
\ee
where
\be 
p={a^2\over\left(R\Omega_0\right)^2}, \quad q={a_A^2\over\left(R\Omega_0\right)^2}, \quad
\bar N^2={N^2\over\Omega_0^2}, \quad g_S={GM\over R^2}, \quad \bar\Omega_z={\Omega_z\over \Omega_0},
\quad \bar\omega={\omega\over \Omega_0},
\quad \Omega_0=\sqrt{GM\over R^3}.
\ee
To solve the dispersion relation for a given neutron star model with $M$ and $R$
and for a given magnetic field $\pmb{B}$, 
we need to supply with appropriate values the following parameters,
\be
p, \quad q, \quad \bar N^2, \quad \bar\Omega_z, \quad \left(Rk\right), \quad \cos\theta, \quad \sin\psi, 
\quad {k_z\over k}, \quad {B_z\over B},
\ee
although we note a relation given by
\be
\cos\theta={k_zB_z+\pmb{k}_H\cdot\pmb{B}_H\over kB}={k_z\over k}{B_z\over B}+{k_H\over k}{B_H\over B}\cos\psi.
\ee
For a given value of $\cos\theta$, we have
\be
\cos\psi={\cos\theta-\left(k_z/k\right)\left(B_z/B\right)\over \left(k_H/k\right)\left(B_H/B\right)},
\ee
and the parameters $k_z/k$ and $B_z/B$ must satisfy an inequality 
$\cos^2\psi\le 1$, that is,
\be
\left({k_z\over k}\right)^2+\left({B_z\over B}\right)^2-2\cos\theta{k_z\over k}{B_z\over B}\le 1-\cos^2\theta,
\ee
which can be rewritten as
\be
{x^2\over x_0^2}+{y^2\over y_0^2}\le1,
\ee
where
\be
x={1\over 2}\left({k_z\over k}+{B_z\over B}\right), \quad 
y={1\over 2}\left({k_z\over k}-{B_z\over B}\right), \quad
x_0^2=\cos^2{\theta\over 2},\quad y_0^2=\sin^2{\theta\over 2}.
\ee
In the following discussions, instead of $k_z/k$ and $B_z/B$, it will be convenient to use the parameters
$f$ and $\theta_f$ defined by
\be
x=x_0f\cos\theta_f \quad {\rm and} \quad y=y_0f\sin\theta_f \quad {\rm with} \quad 0\le f\le 1.
\ee
Using these parameters we have $k_z/k=f\cos\left(\theta_f-\theta/2\right)$ and 
$B_z/B=f\cos\left(\theta_f+\theta/2\right)$, and
$p$, $q$, $\bar N^2$, $\bar\Omega_z$, $\left(Rk\right)$, $\cos\theta$,
$f$, and $\theta_f$ are the parameters we need to specify.

For the fluid ocean of the model NS05T7, typical values of the parameters $p$ and $\bar N^2$
are found to be
\be
p\sim10^{-8}, \quad \bar N^2\sim10^5, 
\ee
and a typical value of the parameter $q$, depending on $B_0$, 
is $q\sim10^{-6}$ for $B_0\sim10^7$G and $q\sim1$
for $B_0\sim10^{10}$G.
Since $k_H/k=\sqrt{1-f^2\cos\left(\theta_f-\theta/2\right)}\ll 1$ for low frequency modes,
we need $f\sim 1$ and $\theta_f\sim\theta/2$.
In the following discussions, for simplicity, we assume $f=0.99$ and $\theta_f=\theta/2$, and $g=g_S$.
Examples of numerical solutions of the dispersion relation (22) are given in Figure 7, where
$\bar\omega$ versus $\bar\Omega_z$ is plotted for $\cos\theta=0.1$ (dotted line), 0.5 (solid line),
and 0.9 (dashed line)
in the left panel, and
$\bar\omega$ versus $\bar\Omega_z$ for 
$q=10^{-8}$ (dotted line), $10^{-6}$ (solid line), and $10^{-4}$ (dashed line)
in the right panel.
It is interesting to note that the asymmetry due to the term proportional to $A_1$ is too weak to become noticeable in the figure.
As shown by the left panel, for a given value of $\cos\theta$, the solution have two branches in this frequency region, 
and the upper and lower branches respectively correspond to gravito-inertial waves, for which
$\omega\propto 2|\Omega_z|$ when $|\Omega_z|$ becomes large, 
and Alfv\'en waves, for which $\omega\propto a_Ak\cos\theta$.
The minimum frequency in the inertial mode branch and the maximum frequency in the Alfv\'en mode branch,
which occur at $\bar\Omega_z\sim0$, increase as $\cos\theta$ increases.
These two branches of modes
are reminiscent of the low frequency waves in the ocean plotted in Figure 1.
We think that the two different mode branches associated with a $mean$ angle $\theta$
between $\pmb{B}$ and $\pmb{k}$ appear as a pair of mode branches in Figure 1 and
that the minimum and maximum frequencies at $\Omega/\sqrt{GM/R^3}\sim0$
in the pair depend on this $mean$ angle, which may vary from one pair to another.
As shown by the right panel of Figure 7, 
for a given value of $\cos\theta$, the frequency of the Alfv\'en modes increases as $q$ increases,
and the inertial branch tends to the relation given by $\omega=2|\Omega_z|$ in the limit of $q\rightarrow 0$.

\begin{figure}
\resizebox{0.5\columnwidth}{!}{
\includegraphics{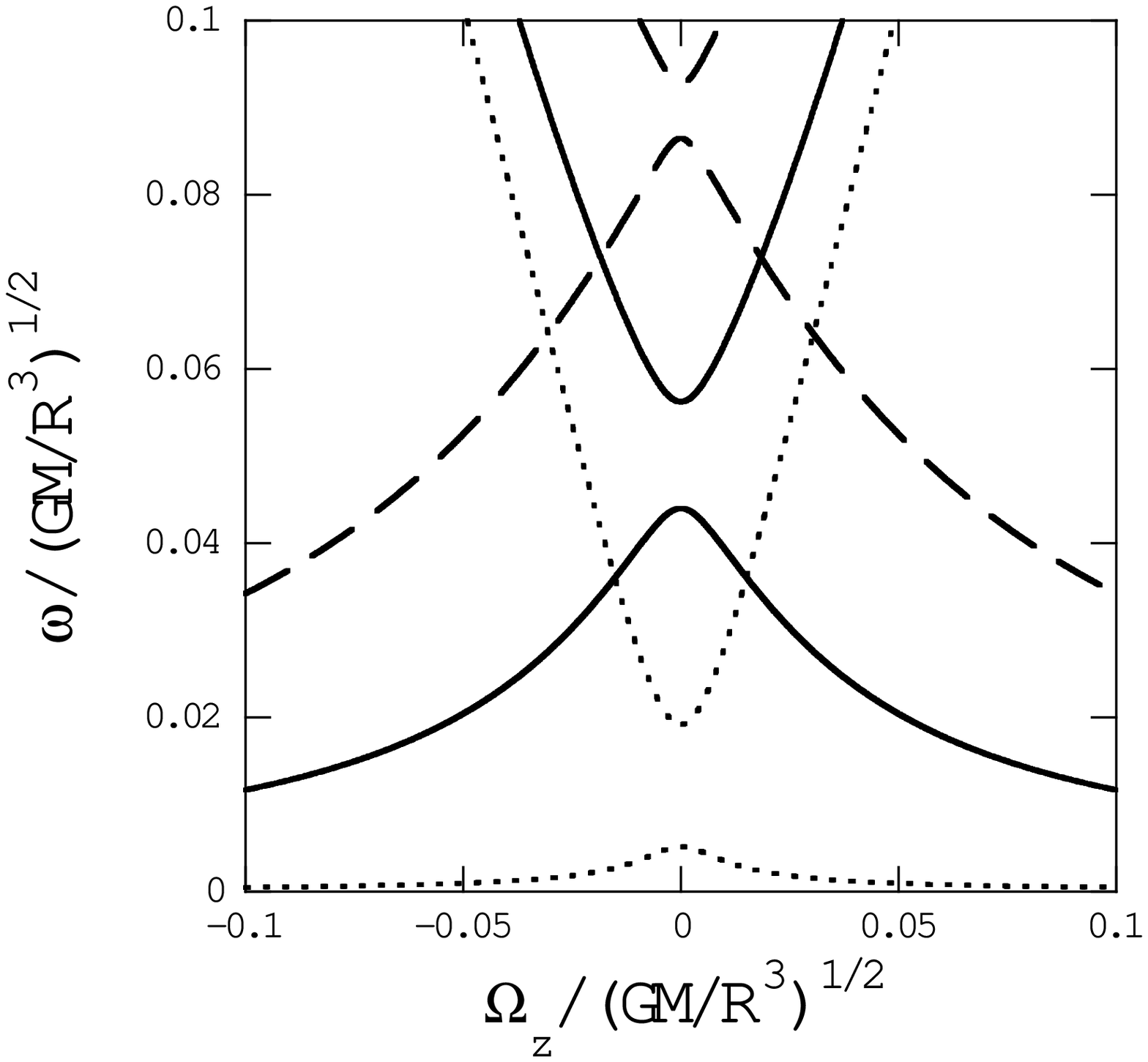}}
\resizebox{0.5\columnwidth}{!}{
\includegraphics{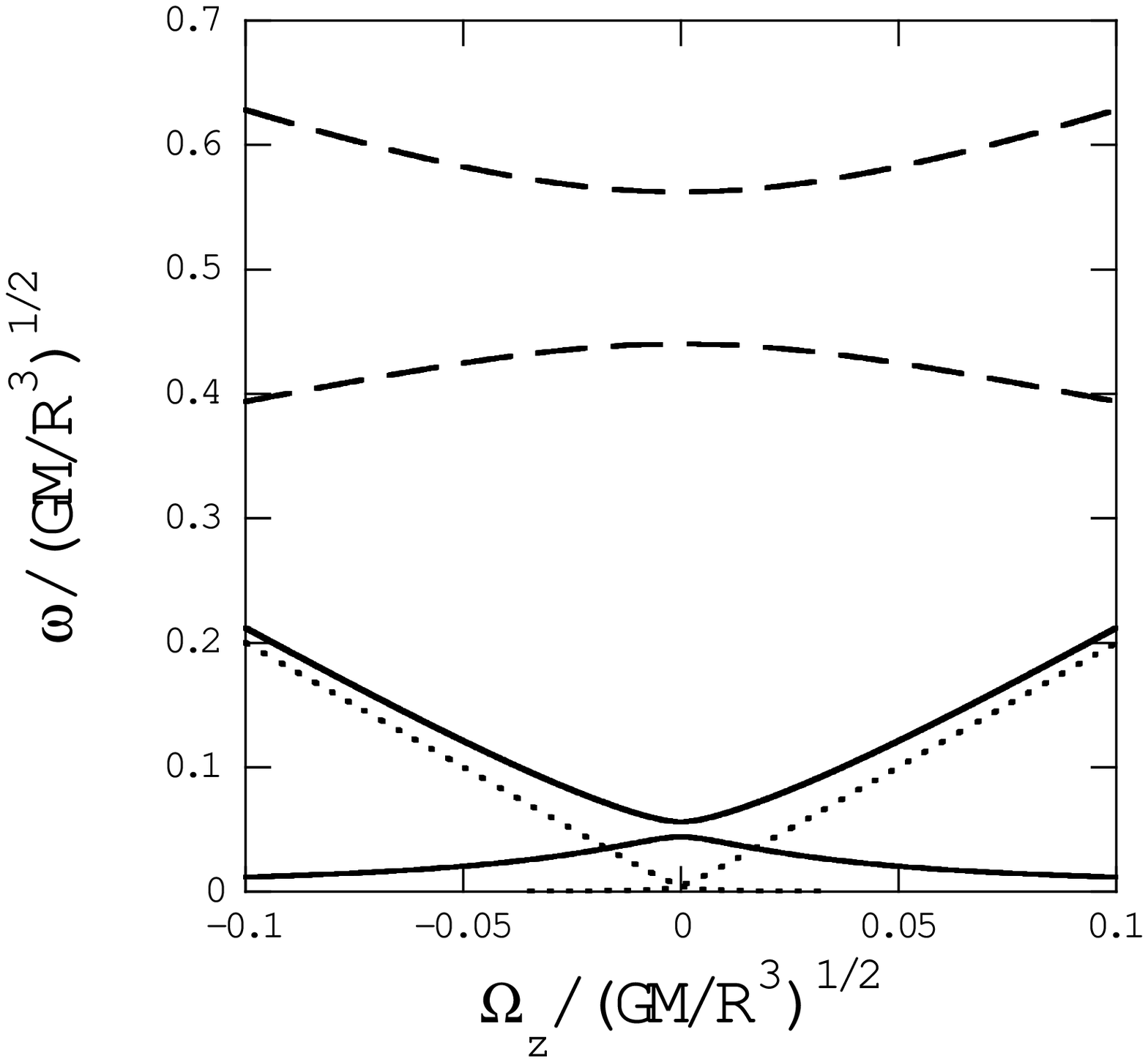}}
\caption{Left panel: Low frequency solutions of the dispersion relation (22) plotted as functions of $\bar\Omega_z$
for $\cos\theta=0.1$ (dotted line), 0.5 (solid line) and 0.9 (dashed line), where
we have assumed $p=10^{-8}$, $q=10^{-6}$, $\bar N^2=10^5$, $(Rk)=10^2$, $f=0.99$, $\theta_f=\theta/2$, and
$g=g_S$.; 
Right panel: Low frequency solutions of the dispersion relation (22) plotted as functions of $\bar\Omega_z$
for $q=10^{-4}$ (dashed line), $10^{-6}$ (solid line), and $10^{-8}$ (dotted line), where
we have assumed $p=10^{-8}$, $\cos\theta=0.5$, $\bar N^2=10^5$, $(Rk)=10^2$, $f=0.99$, $\theta_f=\theta/2$, and
$g=g_S$. 
}
\end{figure}
%

%\section*{Acknowledgments}

\section{discussion and conclusion}

Lamb et al (2009) proposed that the small amplitude, almost sinusoidal 
millisecond X-ray pulsation observed in
accretion powered millisecond X-ray pulsars may be well explained by the hot spot model,
in which the hot spots are assumed to be located at the magnetic poles, which are
nearly aligned with the rotation axis.
As discussed by Lamb et al (2009), even a small drift of the hot spot
could produce appreciable changes in pulsation amplitudes of the X-ray pulsation.
If this proposition is correct, it is interesting to pursue a possibility of 
using the millisecond X-ray pulsation 
to probe the core $r$-modes excited by gravitational wave radiation 
(Andersson 1998; Friedman \& Morsink 1998).
In fact, if the core $r$-modes with $|m|\ge2$ are excited by the emission of gravitational wave,
since the $l^\prime=m=2$ $r$-mode, which is the most strongly destabilized mode among the $r$-modes
(e.g., Lockitch \& Friedman 1999; Yoshida \& Lee 2000a), 
produces the surface displacement vector $\pmb{\xi}$ whose horizontal and toroidal components 
at the surface have large amplitudes around the rotation axis as shown by Figure 6, 
the hot spot could suffer periodic disturbance from the $r$-mode.
Note that the $r$-mode induced temperature perturbation, the surface pattern of which may be proportional to 
$X^r\left(\theta\right)e^{{\rm i}m\phi}$, might generate X-ray variations,
the amplitudes of which should be very small.
If we write the oscillation frequency of $r$-modes as
\be
{\omega/\Omega}=\kappa_0+\kappa_2\bar\Omega^2+O\left(\bar\Omega^4\right),
\ee
the coefficient $\kappa_0$ for the modes is given by
\be
\kappa_0=2m/ \left[l^\prime\left(l^\prime+1\right)\right],
\ee
and the coefficient $\kappa_2$ may depend on the equation of state and the deviation from
the isentropic stratification in the core (e.g., Yoshida \& Lee 2000a,b), 
where $\bar\Omega=\Omega/\Omega_0$.
Since the neutron star core is nearly isentropic such that $N^2\sim0$, we only
have to consider the $l^\prime=|m|$ $r$-modes, for which we have 
$\omega\approx\kappa_0\Omega=2\Omega/\left(|m|+1\right)$, and we obtain
the frequency $\omega\approx 2\Omega/3$ for $m=2$
in the corotating frame of the star and the frequency 
$\sigma\equiv\omega-m\Omega\approx -4\Omega/3$
in an inertial frame.
It may be interesting to point out that
if the $r$-mode of $l^\prime=m=1$ is also excited by some mechanism, this $r$-mode can produce long period variations in an inertial frame since the inertial frame frequency $\sigma\sim0$
for this mode.
Although no detection of periodicities whose frequency is approximately equal to $4\Omega/3$ 
has so far been reported,
if we detect the periodicities produced by the core $r$-modes of $l^\prime=|m|$ in the X-ray millisecond pulsation,
we can use the frequency deviation given by 
$\Delta\bar\omega\equiv\bar\omega-\kappa_0\bar\Omega\approx\kappa_2\bar\Omega^3$ to derive
information about the equations of state and the thermal stratification in the core.
The detectability of the $r$-mode pulsation may depend on the thickness of the crust,
the property of the fluid ocean, and the strength of the magnetic field and so on.

We have calculated non-axisymmetric low frequency modes of a rotating and magnetized neutron star, where we used a neutron star model
composed of a surface fluid ocean, a solid crust, and a fluid core.
We have assumed that the star is threaded by a dipole magnetic field but
the fluid core can be treated as a non-magnetic region.
For this model,
we found that for a magnetic field of strength $B_0\sim10^7$G, Alfv\'en waves in the surface ocean 
come in as low frequency modes, which largely modify the gravito-inertial modes in the ocean.
However, the oscillation frequencies of the modes in the ocean are 
dependent on $j_{\rm max}$, and they do not reach any good convergence 
even if $j_{\rm max}$ is increased to $\sim20$.
At present it is not clear that these ocean modes will converge to discrete modes with real frequencies
in the limit of $j_{\rm max}\rightarrow \infty$.
We also found that no $r$-modes, which are confined to the surface ocean, 
can be found in the presence of the weak magnetic field.
If this is also the case for the accreted envelopes expected in mass accreting
neutron stars in binary systems, we need to reconsider the Rossby wave model for the burst oscillation
in LMXBs (Heyl 2004, Lee 2004).
On the other hand, the toroidal crust modes and the interfacial modes at the core/crust interface,
which show good convergence for $j_{\rm max}\gtsim10$,
are found to be insensitive to magnetic fields of strength $B_0\ltsim10^{12}$G.
We also find that the core $r$-modes and inertial modes are not affected by the magnetic field
even if their eigenfunctions extend to the surface through the magnetized crustal and envelope regions.
However, this will not be the case for neutron stars having magnetic fields as strong as $B_0\sim10^{15}$G
(e.g., Lee 2008).

In the present paper, we employed for modal analysis a low mass neutron star model 
having a thick solid crust and 
a cold and thin surface ocean, for which the toroidal crust modes of low radial order and
low spherical harmonic degree are well separated from the $f$- and $p$-modes.
If we use more massive neutron stars with a hot accreted fluid envelope and a thin solid crust,
the frequencies of the crust modes of low radial order and those of the $f$- and $p$-modes may overlap, 
leading to more complicated frequency spectra.
We think it necessary to conduct similar modal analyses for such neutron star models
to clarify the properties of the ocean modes in the presence of a magnetic field and 
to examine the possibility that the small amplitude millisecond X-ray pulsation
can be used as a probe into the core $r$-modes.

\begin{appendix}

\section{Oscillation equations, jump conditions, and boundary conditions}

In this Appendix, we present the oscillation equations solved for non-axisymmetric 
($m\not=0$) modes of rotating and magnetized neutron stars.
As noted in the text (\S 2), we assume a dipole magnetic field whose axis is aligned with the rotation axis.
We use Newtonian dynamics, and employ the Cowling approximation, neglecting the Euler perturbation of the
gravitational potential.
Using the dependent variables defined by
\be
(\pmb{y}_1)_j=S_{l_j}, \quad (\pmb{y}_2)_j={p^\prime_{l_j}\over gr\rho}, \quad (\pmb{y}_3)_j=H_{l_j}, \quad
(\pmb{y}_4)_j=iT_{l^\prime_j}, \quad (\bmath{b}^H)_j=b^H_{l^\prime_j}, \quad
(\bmath{b}^T)_j=b^T_{l_j}, \quad (\bmath{b}^S)_j=b^S_{l^\prime_j},
\ee
and
\be
\pmb{y}_5=\left(\pmbmt{M}_1\pmb{b}^H+m\pmb{\Lambda}_0^{-1}i\pmb{b}^T\right)/\alpha, 
\quad \pmb{y}_6=\left(m\pmb{\Lambda}_1^{-1}\pmb{b}^H+\pmbmt{M}_0i\pmb{b}^T\right)/\alpha,
\ee
the oscillation equation for non-axisymmetric modes with $m\not=0$ for fluid regions threaded by 
a dipole magnetic field is given by 
\be
r{d\pmb{y}_1\over dr}=\left({V\over\Gamma_1}-3\right)\pmb{y}_1-{V\over\Gamma_1}\pmb{y}_2+\pmb{\Lambda}_0\pmb{y}_3,
\ee
\be
r{d\pmb{y}_2\over dr}=\left(rA+c_1\bar\omega^2\right)\pmb{y}_1
+\left(1-rA-U\right)\pmb{y}_2-m\nu c_1\bar\omega^2\pmb{y}_3-\nu c_1\bar\omega^2\pmbmt{C}_0\pmb{y}_4
+{c_1\bar\omega^2\over2}\pmb{R},
\ee
\be
\pmbmt{M}_0r{d\pmb{y}_3\over dr}+m\pmb{\Lambda}_1^{-1}r{d\pmb{y}_4\over dr}=-{1\over 2}\left({V\over\Gamma_1}-4\right)\pmbmt{K}\pmb{y}_1
+{1\over 2}{V\over\Gamma_1}\pmbmt{K}\pmb{y}_2+\left(\pmbmt{M}_0-{1\over2}\pmbmt{C}_1\right)\pmb{y}_3+m\pmb{\Lambda}_1^{-1}\pmb{y}_4
-{1\over2}\alpha{\pmb{b}^H\over\alpha},
\ee
\be
m\pmb{\Lambda}_0^{-1}r{d\pmb{y}_3\over dr}+\pmbmt{M}_1r{d\pmb{y}_4\over dr}={1\over 2}\left({V\over\Gamma_1}-4\right)m\pmb{\Lambda}_0^{-1}\pmb{y}_1
-{1\over 2}{V\over\Gamma_1}m\pmb{\Lambda}_0^{-1}\pmb{y}_2+m\pmb{\Lambda}_0^{-1}\pmb{y}_3+\left(\pmbmt{M}_1-{1\over 2}\pmbmt{C}_0\right)\pmb{y}_4
-{1\over 2}\alpha{i\pmb{b}^T\over\alpha},
\ee
\be
r{d\pmb{y}_5\over dr}=
\left(-m\nu \pmb{\Lambda}_0^{-1}+{\pmbmt{M}_1\pmbmt{C}_0^T\over\alpha}\right)\pmb{y}_1-{\pmb{y}_2\over c_1\bar\omega^2}
+\left(\pmbmt{L}_0-{2\pmbmt{M}_1\pmb{\Lambda}_1\pmbmt{M}_0\over\alpha}\right)\pmb{y}_3
-\left(\nu +{2m\over\alpha}\right)\pmbmt{M}_1\pmb{y}_4 +\left(2-{d\ln\alpha\over d\ln r}\right)\pmb{y}_5
+{m\over2}{i\pmb{b}^T\over\alpha},
\ee
\be
r{d\pmb{y}_6\over dr}=\left(\nu -{m\over\alpha}\right)\pmbmt{K}\pmb{y}_1
-\left(\nu +{2m\over\alpha}\right)\pmbmt{M}_0\pmb{y}_3
+\left(\pmbmt{L}_1-{2m^2\over\alpha}\pmb{\Lambda}_1^{-1}\right)\pmb{y}_4
+\left(2-{d\ln\alpha\over d\ln r}\right)\pmb{y}_6-{1\over 2}\pmbmt{C}_1{i\pmb{b}^T\over\alpha},
\ee
with
\be
\pmb{R}=r{d\over dr}\left(\pmbmt{C}_0{\pmb{b}^H\over\alpha}+m{i\pmb{b}^T\over \alpha}\right)
-\left(2-{d\ln\alpha\over d\ln r}\right)\left(\pmbmt{C}_0{\pmb{b}^H\over\alpha}+m{i\pmb{b}^T\over \alpha}\right)
-\pmbmt{C}_0{\pmb{b}^S\over\alpha}
\ee
and
\be
\bmath{b}^S=-\pmb{\Lambda}_1\pmbmt{K}\bmath{y}_1-2\pmb{\Lambda}_1 \pmbmt{M}_0\bmath{y}_3-2m\bmath{y}_4,
\ee
where $\nu=2\Omega/\omega$, and
$\bar\omega=\omega/\sqrt{GM/R^3}$ with
$M$ and $R$ being the mass and radius of the star, and
\be
U={d\ln M_r\over d\ln r}, \quad V=-{d\ln p\over d\ln r}, \quad
c_1={(r/R)^3\over M_r/M}, \quad
\alpha={c_1\bar\omega^2p V\over 4p_{\rm B}}, \quad p_{\rm B}={B_0^2\over 8\pi},
\ee
and $\pmbmt{C}_0^T$ is the transposed matrix of $\pmbmt{C}_0$.
The non-zero elements of the matrices $\pmbmt{M}_0$, $\pmbmt{M}_1$, $\pmbmt{C}_0$, $\pmbmt{C}_1$, $\pmbmt{K}$, $\pmb{\Lambda}_0$, and $\pmb{\Lambda}_1$ 
that appear in the oscillation equation given above are defined by
\be
(\pmbmt{M}_0)_{j,j}={l_j\over l_j+1}J^m_{l_j+1}, \quad
(\pmbmt{M}_0)_{j,j+1}={l_j+3\over l_j+2}J^m_{l_j+2}, \quad
(\pmbmt{M}_1)_{j,j}={l_j+2\over l_j+1}J^m_{l_j+1}, \quad
(\pmbmt{M}_1)_{j+1,j}={l_j+1\over l_j+2}J^m_{l_j+2},
\ee
\be
(\pmbmt{C}_0)_{j,j}=-(l_j+2)J^m_{l_j+1}, \quad (\pmbmt{C}_0)_{j+1,j}=(l_j+1)J^m_{l_j+2}, \quad
(\pmbmt{C}_1)_{j,j}=l_jJ^m_{l_j+1}, \quad (\pmbmt{C}_1)_{j,j+1}=-(l_j+3)J^m_{l_j+2},
\ee
\be
(\pmbmt{K})_{j,j}={J^m_{l_j+1}\over l_j+1}, \quad (\pmbmt{K})_{j,j+1}=-{J^m_{l_j+2}\over l_j+2}, 
\quad (\pmb{\Lambda}_0)_{j,j}=l_j\left(l_j+1\right), \quad (\pmb{\Lambda}_1)_{j,j}=l^\prime_j\left(l^\prime_j+1\right)
\ee
for even modes, and
\be
(\pmbmt{M}_0)_{j,j}={l_j+1\over l_j}J^m_{l_j}, \quad
(\pmbmt{M}_0)_{j+1,j}={l_j\over l_j+1}J^m_{l_j+1}, \quad
(\pmbmt{M}_1)_{j,j}={l_j-1\over l_j}J^m_{l_j}, \quad
(\pmbmt{M}_1)_{j,j+1}={l_j+2\over l_j+1}J^m_{l_j+1},
\ee
\be
(\pmbmt{C}_0)_{j,j}=(l_j-1)J^m_{l_j}, \quad (\pmbmt{C}_0)_{j,j+1}=-(l_j+2)J^m_{l_j+1}, \quad
(\pmbmt{C}_1)_{j,j}=-(l_j+1)J^m_{l_j}, \quad (\pmbmt{C}_1)_{j+1,j}=l_jJ^m_{l_j+1},
\ee
\be
(\pmbmt{K})_{j,j}=-{J^m_{l_j}\over l_j}, \quad (\pmbmt{K})_{j+1,j}={J^m_{l_j+1}\over l_j+1}, 
\quad (\pmb{\Lambda}_0)_{j,j}=l_j\left(l_j+1\right), \quad (\pmb{\Lambda}_1)_{j,j}=l^\prime_j\left(l^\prime_j+1\right)
\ee
for odd modes, where
\be
J_l^m=\sqrt{l^2-m^2\over 4l^2-1}
\ee
for $l\ge|m|$, and $J_l^m=0$ for $l<|m|$.
The matrices $\pmbmt{L}_0$ and $\pmbmt{L}_1$ are given by
\be
\pmbmt{L}_0=\pmbmt{1}-m\nu\pmb{\Lambda}_0^{-1}, \quad \pmbmt{L}_1=\pmbmt{1}-m\nu\pmb{\Lambda}_1^{-1}
\ee 
with $\bmath{1}$ being the unit matrix.
Note that $l_j=|m|+2(j-1)$ and $l^\prime_j=l_j+1$ for even modes, 
and $l_j=|m|+2j-1$ and $l^\prime_j=l_j-1$ for odd modes.

For a solid region threaded by the dipole magnetic field, on the other hand, we use the dependent variables defined as
\be
(\pmb{z}_1)_j=S_{l_j}, \quad (\pmb{z}_2)_j=H_{l_j}, \quad
(\pmb{z}_3)_j=iT_{l^\prime_j}, 
\ee
and
\be
\pmb{z}_4=\left(\Gamma_1-{2\over 3}\alpha_1\right)
\left[{1\over r^2}{d\over dr}\left(r^3\pmb{z}_1\right)-\pmb{\Lambda}_0\pmb{z}_2\right]
+2\alpha_1{d\over dr}\left(r\pmb{z}_1\right)
+{2p_{\rm B}\over p}\left(\pmbmt{C}_0\pmb{b}^H+mi\pmb{b}^T\right),
\ee
\be
\pmb{z}_5=\alpha_1\left(r{d\pmb{z}_2\over dr}+\pmb{z}_1\right)
-{4p_{\rm B}\over p}\left(\pmbmt{M}_1\pmb{b}^H+m\pmb{\Lambda}_0^{-1}i\pmb{b}^T\right), \quad
\pmb{z}_6=\alpha_1r{d\pmb{z}_3\over dr}-{4p_{\rm B}\over p}\left(m\pmb{\Lambda}_1^{-1}\pmb{b}^H+\pmbmt{M}_0i\pmb{b}^T\right),
\ee
where 
\be
\alpha_1={\mu\over p}, \quad \alpha_2=\Gamma_1-{2\over 3}\alpha_1, 
\quad \alpha_3=\Gamma_1+{4\over 3}\alpha_1.
\ee
The oscillation equation then becomes
\be
r{d\pmb{z}_1\over dr}=-{3\Gamma_1\over\alpha_3}\pmb{z}_1+{\alpha_2\over\alpha_3}\pmb{\Lambda}_0\pmb{z}_2
+{1\over\alpha_3}\pmb{z}_4-{2p_{\rm B}\over \alpha_3p}\left(\pmbmt{C}_0\pmb{b}^H+mi\pmb{b}^T\right),
\ee
\begin{eqnarray}
\pmbmt{M}_0r{d\pmb{z}_2\over dr}+m\pmb{\Lambda}_1^{-1}r{d\pmb{z}_3\over dr}=
{1\over 2}\left(1+{3\Gamma_1\over\alpha_3}\right)\pmbmt{K}\pmb{z}_1
+\left(\pmbmt{M}_0-{\alpha_2\over2\alpha_3}\pmbmt{C}_1\right)\pmb{z}_2
+m\pmb{\Lambda}_1^{-1}\pmb{z}_3-{1\over 2\alpha_3}\pmbmt{K}\pmb{z}_4 \nonumber \\
+\left({p_{\rm B}\over \alpha_3 p}\pmbmt{K}\pmbmt{C}_0-{1\over 2}\pmbmt{1}\right)\pmb{b}^H
+{p_{\rm B}\over \alpha_3 p}m\pmbmt{K}i\pmb{b}^T,
\end{eqnarray}
\begin{eqnarray}
m\pmb{\Lambda} _0 ^{ - 1} r{d\pmb{z}_2\over dr } + \pmbmt{M}_1 r{d\pmb{z}_3\over dr} = 
  - {1\over 2}\left( 1 + {3\Gamma _1 \over \alpha _3 } \right)m\pmb{\Lambda} _0 ^{ - 1} \pmb{z}_1  
+ m\left[ {\pmb{\Lambda} _0 ^{ - 1}  
+ {1\over 2}\left( {\alpha _2 \over \alpha _3 } - 1 \right)\pmbmt{1}} \right]\pmb{z}_2  
+ \left( \pmbmt{M}_1  - {1\over 2}\pmbmt{C}_0  \right)\pmb{z}_3  
+ {m\over 2\alpha _3 }\pmb{\Lambda} _0 ^{ - 1} \pmb{z}_4  \nonumber \\ 
- {p_{\rm B} \over \alpha _3 p}m\pmb{\Lambda} _0 ^{ - 1} \pmbmt{C}_0 \pmb{b}^H  
- {1\over 2}\left( \pmbmt{1} + {2p_{\rm B} \over \alpha _3 p}m^2 \pmb{\Lambda} _0 ^{ - 1}  \right)i\pmb{b}^T,
\end{eqnarray}
\begin{eqnarray}
r{d\pmb{z}_4\over dr}=\left[\left(UV-4V-c_1\bar\omega^2V+{12\alpha_1\Gamma_1\over\alpha_3}
\right)\pmbmt{1}-{2p_{\rm B}\over p}\pmbmt{C}_0\pmb{\Lambda}_1 \pmbmt{K}\right]\pmb{z}_1 
+\left[m c_1\bar\omega^2V\nu\pmbmt{1}+\left(V-2\alpha_1-{4\alpha_1\alpha_2\over\alpha_3}\right)\pmb{\Lambda}_0-{4p_{\rm B}\over p}\pmbmt{C}_0\pmb{\Lambda}_1\pmbmt{M}_0
\right]\pmb{z}_2  \nonumber \\ 
+\left(c_1\bar\omega^2V\nu-{4p_{\rm B}\over p}m\right)\pmbmt{C}_0\pmb{z}_3+\left(V-{4\alpha_1\over\alpha_3}\right)\pmb{z}_4+\pmb{\Lambda}_0\pmb{z}_5  
+{4p_{\rm B}\over p}\left[\pmb{\Lambda}_0\pmbmt{M}_1-2\left(1-{\alpha_1\over\alpha_3}\right)\pmbmt{C}_0\right]\pmb{b}^H
-{4p_{\rm B}\over p}\left(1-{2\alpha_1\over\alpha_3}\right)mi\pmb{b}^T,
\end{eqnarray}
\begin{eqnarray}
r{d\pmb{z}_5\over dr}=\left[\left(V
-{6\alpha_1\Gamma_1\over\alpha_3}\right)\pmbmt{1}+mc_1\bar\omega^2V\nu\pmb{\Lambda}_0^{-1}
+{4p_{\rm B}\over p}\pmbmt{M}_1\pmb{\Lambda}_1\pmbmt{K}\right]\pmb{z}_1
+\left[-c_1\bar\omega^2V\pmbmt{L}_0
+{8p_{\rm B}\over p}\pmbmt{M}_1\pmb{\Lambda}_1\pmbmt{M}_0
-2\alpha_1\pmbmt{1}+2\alpha_1\left(1+{\alpha_2\over\alpha_3}\right)
\pmb{\Lambda}_0\right]\pmb{z}_2\nonumber \\ 
+\left(c_1\bar\omega^2V\nu+{8mp_{\rm B}\over p}\right)\pmbmt{M}_1\pmb{z}_3-\left(1-{2\alpha_1\over\alpha_3}\right)\pmb{z}_4
+\left(V-3\right)\pmb{z}_5+{4p_{\rm B}\over p}\left[\pmbmt{M}_1+\left({1\over 2}-{\alpha_1\over\alpha_3}\right)\pmbmt{C}_0\right]\pmb{b}^H
+{4p_{\rm B}\over p}m\left(\pmb{\Lambda}_0^{-1}-{\alpha_1\over\alpha_3}\pmbmt{1}\right)i\pmb{b}^T,
\end{eqnarray}
\begin{eqnarray}
r{d\pmb{z}_6\over dr}=\left(-c_1\bar\omega^2V\nu+{4p_{\rm B}\over p}m\right)\pmbmt{K}\pmb{z}_1
+\left(c_1\bar\omega^2V\nu+{8p_{\rm B}\over p}m\right)\pmbmt{M}_0\pmb{z}_2
+\left[-c_1\bar\omega^2V\pmbmt{L}_1+{8p_{\rm B}\over p}m^2\pmb{\Lambda}_1^{-1}
-2\alpha_1\left(\pmbmt{1}-{1\over 2}\pmb{\Lambda}_1\right)\right]\pmb{z}_3 \nonumber\\
+\left(V-3\right)\pmb{z}_6
+{4p_{\rm B}\over p}m\pmb{\Lambda}_1^{-1}\pmb{b}^H
+{4p_{\rm B}\over p}\left(\pmbmt{M}_0+{1\over 2}\pmbmt{C}_1\right)i\pmb{b}^T.
\end{eqnarray}

We derive one set of the jump conditions imposed at the interface between the solid crust and the fluid ocean 
by assuming the continuity condition of the displacement vector at the interface (see e.g., Lee 2007):
\be
\left[\pmb{\xi}\left(r_i\right)\right]^+_-=0,
\ee
where $\left[F\left(r_i\right)\right]^+_-
\equiv\lim_{\epsilon\rightarrow 0}\left[F(r_i+\epsilon)-F(r_i-\epsilon)\right]$.
The other set of the jump conditions are derived 
from the continuity condition of the $rr$, $r\theta$, and $r\phi$
components of the perturbed traction at the interface;
\be
\left[\delta\tau_{rj}\left(r_i\right)\right]^+_-=0,
\ee
where $\delta\tau_{ij}$ denotes the $ij$ component of the perturbed traction, and the 
detail expression of $\delta\tau_{ij}$ is given in Lee (2007).
As for the jump conditions at the interface between the solid crust and the fluid core, which is
assumed non-magnetic, we use the continuity of the radial component of the displacement vector
and of the  $rr$, $r\theta$, and $r\phi$
components of the perturbed traction at the interface.

The surface boundary conditions for non-axisymmetric modes are given by
\be
\delta p/p=0,\quad i\pmb{b}^T=0, \quad \pmb{b}^S+L^+\pmb{b}^H=0,
\ee
where $\delta p$ denotes the lagrangian perturbation of the pressure, and $L^+=\delta_{ij}\left(l^\prime_j+1\right)$  (Lee 2007).
As for the inner boundary conditions at stellar center, we require that the functions $r\pmb{y}_1$
and $r\pmb{y}_2$ are regular at the center.

\end{appendix}

\end{document}